\documentclass[preprint]{aastex}
\usepackage{amsmath}
\usepackage{amssymb}
\usepackage{graphicx}
\usepackage{bm}
\usepackage{natbib}
\usepackage{longtable}

\begin{document}
\title{Three-Dimensional MHD simulation of Caltech Plasma Jet Experiment: First results}

\author{Xiang Zhai}
\affil{Applied Physics, California Institute of Technology, Pasadena, CA 91125}
\email{xzhai@caltech.edu}
\author{Hui Li}
\affil{Theoretical Division, Los Alamos National Laboratory, Mail Stop B227, Los Alamos, NM 87545}
\email{hli@lanl.gov}
\author{Paul M. Bellan}
\affil{Applied Physics, California Institute of Technology, Pasadena, California 91125}
\email{pbellan@caltech.edu}
\author{Shengtai Li}
\affil{Mathematical Modeling and Analysis, Los Alamos National Laboratory, Mail Stop B284, Los Alamos, NM 87545}
\email{sli@lanl.gov}

\shorttitle{MHD simulation of Caltech jet}
\shortauthors{Zhai et al.}

\date{\today}

\begin{abstract}

Magnetic fields are believed to play an essential role in astrophysical jets with observations suggesting the presence of helical magnetic fields. Here, we present three-dimensional (3D) ideal MHD simulationsof the Caltech plasma jet experiment using a magnetic tower scenario as the baseline model. Magnetic fields consist of an initially localized dipole-like poloidal component and a toroidal component that is continuously being injected into the domain. This flux injection mimics the poloidal currents driven by the anode-cathode voltage drop in the experiment. The injected toroidal field stretches the poloidal fields to large distances, while forming a collimated jet along with several other key features. Detailed comparisons between 3D MHD simulations and experimental measurements provide a comprehensive description of the interplay among magnetic force, pressure and flow effects. In particular, we delineate both the jet structure and the transition process that converts the injected magnetic energy to other forms.  With suitably chosen parameters that are derived from experiments, the jet in the simulation agrees quantitatively with the experimental jet in terms of magnetic/kinetic/inertial energy, total poloidal current, voltage, jet radius, and jet propagation velocity. Specifically, the jet velocity in the simulation is proportional to the poloidal current divided by the square root of the jet density, in agreement with both the experiment and analytical theory. This work provides a new and quantitative method for relating experiments, numerical simulations and astrophysical observation, and demonstrates the possibility of using terrestrial laboratory experiments to study astrophysical jets.

\end{abstract}


\maketitle

\section{Introduction} 

Magnetohydrodynamic (MHD) plasma jets exist in a wide variety of systems from 
terrestrial experiments to astrophysical objects, and have attracted substantial 
attention for decades. For example, energetic and usually relativistic jets are 
commonly observed originating from active galactic nuclei  (AGNs), which are believed to be
powered by supermassive black holes. AGN jets usually remain highly collimated 
for tens to hundreds of kiloparsecs from the host galaxy core 
\citep[e.g.,][]{Ferrari_1998}.
It is generally accepted that AGN jets are powered by the central black 
hole accretion disk region. On a much smaller scale, stellar jets are believed to 
be an integral part of star formation with an active accretion disk surrounding a 
young star \citep[e.g.,][]{Hartigan_Hillenbrand_2009}.

Despite our limited understanding of how the disks or central objects 
produce collimated jets, observational evidence has shown that magnetic 
fields are crucial in collimating and accelerating jets. Highly polarized 
synchrotron radiation is observed from both AGN jets and stellar jets, 
implying that jets have a strongly organized magnetic field. 
For example, the two lobes of T Tauri S, created by the interaction of 
a bipolar stellar jet with the remote interstellar medium (ISM), exhibit 
strong circularly polarized radio emission with opposite helicity \citep{Ray_et_al_1997}. 
Large-scale magnetic fields from bipolar AGN jets also show 
transverse asymmetries \citep{Clausen-Brown_2011}. These 
observations strongly suggest that a large-scale poloidal magnetic field, 
centered at the accretion disk or the central object, plays a crucial role in 
generating and propagating both AGN jets and stellar jets.
A close look into the jet origin of M87 has found that the jet at $100$ 
Schwarzschild radii is only weakly collimated (opening angle $\approx 60^{\circ}$), 
but becomes very collimated at larger distance (opening angle $< 10^{\circ}$). 
This favors a magnetic collimation mechanism at $z>100$ Schwarzschild 
radii \citep{Junor_et_al_1999}. The 3C31 jet and several other AGN jets
exhibit a global kink-like $m=1$ instability or helical wiggles 
\citep{Hardee_2008, Nakamura_2007}, implying the existence of a 
strong axial current along the jet, or, equivalently, a strong toroidal 
magnetic field around the jet. 
Here, we define the central axis along the jet as the $z$ axis of a cylindrical coordinate system. 
The $r$ and $z$ directions are called the ``poloidal" direction and the 
azimuthal direction $\theta$ is called the ``toroidal" direction.
These facts suggest a $z$-pinch type of collimation mechanism, in which the 
axial current $J_z$ and the associated azimuthal magnetic field $B_{\theta}$ 
generate a radial Lorentz force and squeeze the jet plasma against the 
pressure gradient at the central region of the jet.

The surprising similarities of astrophysical jets in morphology, kinetic behavior 
and magnetic field configuration over vastly different scales have inspired many 
efforts to model these jets using ideal MHD theory. 
One important feature is that ideal MHD theory has no intrinsic scale. 
Therefore an ideal MHD model is highly scalable and capable of describing a range of 
systems having many orders of magnitude difference in size. Ideal MHD theory 
assumes that the Lundquist number, a dimensionless measurement of plasma 
conductivity, to be infinite. This leads to the well-known ``frozen-in" condition, 
wherein magnetic flux is frozen into the plasma and moves together with 
the plasma \citep{Bellan_Plasma}. Hence the evolution of plasma material 
and magnetic field configuration is unified in ideal MHD. \citet{Blandford_1982} 
developed a self-similar MHD model, in which a magnetocentrifugal mechanism 
accelerates plasma along poloidal field lines threading the accretion disk; 
the plasma is then collimated by a toroidal dominant magnetic field at 
larger distance. \citet{Lynden-Bell_1996, Lynden-Bell_2003} and \citet{SherwinLyndenBell_2007} constructed an analytical magnetostatic MHD model where the upward flux of a dipole magnetic field is twisted 
relative to the downward flux. The height of the magnetically dominant 
cylindrical plasma grows in this configuration. The toroidal component of the 
twisted field is responsible for both collimation and propagation. 
The \citet{Lynden-Bell_1996, Lynden-Bell_2003} and \citet{SherwinLyndenBell_2007} model and various following models, 
(typically numerical simulations with topologically similar magnetic field configurations; 
e.g., \citet{Li_2001,Lovelace_2002,Li2006_MagneticTower,Nakamura_2008,XuLi2008_XrayCavity}), 
are called ``magnetic tower" models. In these models, the large scale magnetic fields
are often assumed to possess ``closed'' field lines with both footprints residing in the disk. 
Because plasma at different radii on the accretion disk and in the corona have different angular velocity, the poloidal magnetic field lines threading the disk 
will become twisted up \citep{Blandford_1982, Lynden-Bell_1996, Lynden-Bell_2003, SherwinLyndenBell_2007, Li_2001, Lovelace_2002},
giving rise to the twist/helicity or the toroidal component of the magnetic fields in the jet. Faraday rotation measurements to 3C 273 show a helical magnetic field structure and an increasing pitch angle between toroidal and poloidal component along the jet \citep{Zavala_Taylor_2005}. These results favor a magnetic structure suggested by magnetic tower models. Furthermore, it is (often implicitly) assumed that the mass loading onto these magnetic 
fields is small, so the communication by Alfv\'en waves is often fast 
compared to plasma flows.

These models have achieved various degrees of success and  
have improved understanding of astrophysical jets significantly. 
However, the limitations of astrophysical observation, e.g., 
mostly unresolved spatial features, 
passive observation and impossibility of in-situ measurement, 
have imposed a natural limitation to these models. During the last decade, 
on the other hand, it has been realized that laboratory experiments can 
provide valuable insights for studying astrophysical jets. 
Laboratory experiments have the intrinsic value of elucidating key physical 
processes (especially those involving magnetic fields) in highly nonlinear systems. 
The relevance of laboratory experiments relies on the scalability of the MHD theory and the 
equivalence of differential rotation of the astrophysical accretion disk to 
voltage difference across the laboratory electrodes (at least in the 
magnetically dominated limit). 
The latter can be seen by considering Ohm's Law in ideal MHD theory, 
$\mathbf{E}+\mathbf{v}\times\mathbf{B}=0$; $\mathbf{E}$ is the electric field 
and $\mathbf{v}$ is the plasma velocity. The radial component of Ohm's Law 
is $E_r+v_zB_{\theta}-v_{\theta}B_z=0$. If we ignore the vertical motion $v_z$ 
of the accretion disk, it is seen that $E_r=v_{\theta}B_z$, i.e., 
an equivalent radial electric field is created by $\theta$ motion (rotation), 
and spatial integration of this electric field gives the voltage difference at 
different radii. Such a voltage difference is relatively easy to create in lab 
experiments by applying a voltage across a coaxial electrode pair
(See Section 3.3 for the discussion on the helicity). In addition, 
experimental jets are reproducible, parameterizable and in-situ measurable. 
They automatically ``calculate" the MHD equations and also ``incorporate" 
non-ideal MHD plasma effects. 
Most importantly, the very 
fact that jets can be produced in the experiments strongly suggests there 
should be relatively simple unifying MHD concept characterizing AGN jets, 
stellar jets and experimental jets \citep{Bellan_2009}.

The experiments carried out at Caltech and Imperial College have used 
pulse-power facilities to simulate ``magnetic tower" astrophysical 
jets \citep[e.g.,][]{Hsu_Bellan_2002, Kumar_Bellan_2009, Lebedev_2005, Ciardi_2007, Ciardi_2009}. 
The two experiments have topologically similar toroidal magnetic field 
configurations and plasma collimation mechanisms. However, in addition to 
the toroidal field, the Caltech plasma jets also have a poloidal magnetic field 
threading a co-planar coaxial plasma gun so the global field configuration 
is possibly more like a real astrophysical situation. Magnetically driven 
jets are produced by both groups, and the jets are collimated and accelerated 
in essentially the manner described by the magnetic tower models. 
Due to the lack of poloidal magnetic field, the plasma jets in the group at Imperial College
undergo violent instability and break into episodic parts (magnetic bubbles). 
The Caltech jets remain very collimated and straight and undergo a kink instability 
when the jet length satisfies the classic Kruskal-Shafranov threshold 
\citep{Hsu_Bellan_2003,Hsu_Bellan_2005}. The Alfv\'{e}nic and supersonic jets 
created by the Caltech group have relatively low thermal to magnetic pressure ratio $\beta \sim 0.1$ and large 
Lundquist number $S \sim 10-100$. Other features including flux rope merging, 
magnetic reconnection, Rayleigh-Taylor instability and jet-ambient gas interaction 
are also produced \citep{Hsu_Bellan_2003, Hsu_Bellan_2005, Yun_Bellan_2010,
Yun_You_Bellan_2007,Moser_Nature, Moser_Bellan_2012}. 
A detailed introduction to the Caltech jet experiments is given in Section 2.

Observation, analytical modeling, numerical simulation and terrestrial 
experiments (laboratory astrophysics) are all crucial approaches for a better understanding 
of astrophysical jets. Compared to observation or analytical models, 
numerical simulation and terrestrial experiments share certain common features, 
such as the ability to deal with more complex structures and sophisticated behaviors, 
larger freedom in the parameter space compared to observation, and more resolution. 
However, cross-validation between numerical simulations and experiments has been 
very limited. Lab experiments can provide detailed validation for numerical models, 
while the numerical models can test the similarity between the terrestrial experimental 
jets and astrophysical jets.

We report here 3D ideal MHD numerical studies that simulate the Caltech plasma 
jet experiment. The numerical model uses a modified version of a computational code 
\citep{LiLi2003_AMR3D} previously given by \citet{Li2006_MagneticTower} 
for simulating AGN jets in the intra-cluster medium. Motivated by both observations 
and experiments, we adopt the approach that 
the jet has a global magnetic field structure and 
both poloidal and toroidal magnetic fields in the simulation are totally 
contained in a bounded volume.
Following the approach in \citet{Li2006_MagneticTower}, 
the MHD equations are normalized to suit the experiment scale. An initial poloidal field 
configuration is chosen to simulate the experimental bias field configuration and 
the toroidal magnetic field injection takes a compact form to represent the electrodes. 
Detailed comparisons between simulation and experiment have been undertaken, 
addressing the collimation and acceleration mechanism, jet morphology, 
axial profiles of density and magnetic field and the 3D magnetic field structure. 
The simulations have reproduced most salient features of the experimental jet 
quantitatively, with discrepancies generally less than a factor of three for key quantities. 
The conversion of magnetic to kinetic energy from jet base to jet head is 
examined in the simulation and compared to the experiment. As a result, 
a Bernoulli-like equation, stating that the sum of kinetic and toroidal magnetic 
field energy is constant along the axial extent of the jet, is validated by 
analytical modeling, laboratory experiment and the numerical simulation.

The paper is organized as follows: In Section 2, we introduce the Caltech plasma jet 
experiment and demonstrate that the global behavior of the experimental jets 
can be described by ideal MHD theory. In Section 3 we describe the approach and 
configuration of our simulation, and show that the compact toroidal magnetic 
field injection method used in the simulation is equivalent to the energy and 
helicity injection through the electrodes used in the experiment. In Section 4.1, 
we present the simulation results of a typical run, and compare these results 
with experimental measurements. In Section 4.2, we perform multiple simulations 
with different toroidal injection rates and examine the jet velocity dependence 
on poloidal current. These results together with experimental measurements 
confirm the MHD Bernoulli equation and the magnetic to kinetic energy 
conversion in the MHD driven plasma jet. Section 5 discusses the sensitivity of the simulation results to initial and injection conditions. Conclusions and discussions are given in Section 6.

\section{Caltech plasma jet experiment}

\begin{figure}
\figurenum{1}
\epsscale{0.8}
\plotone{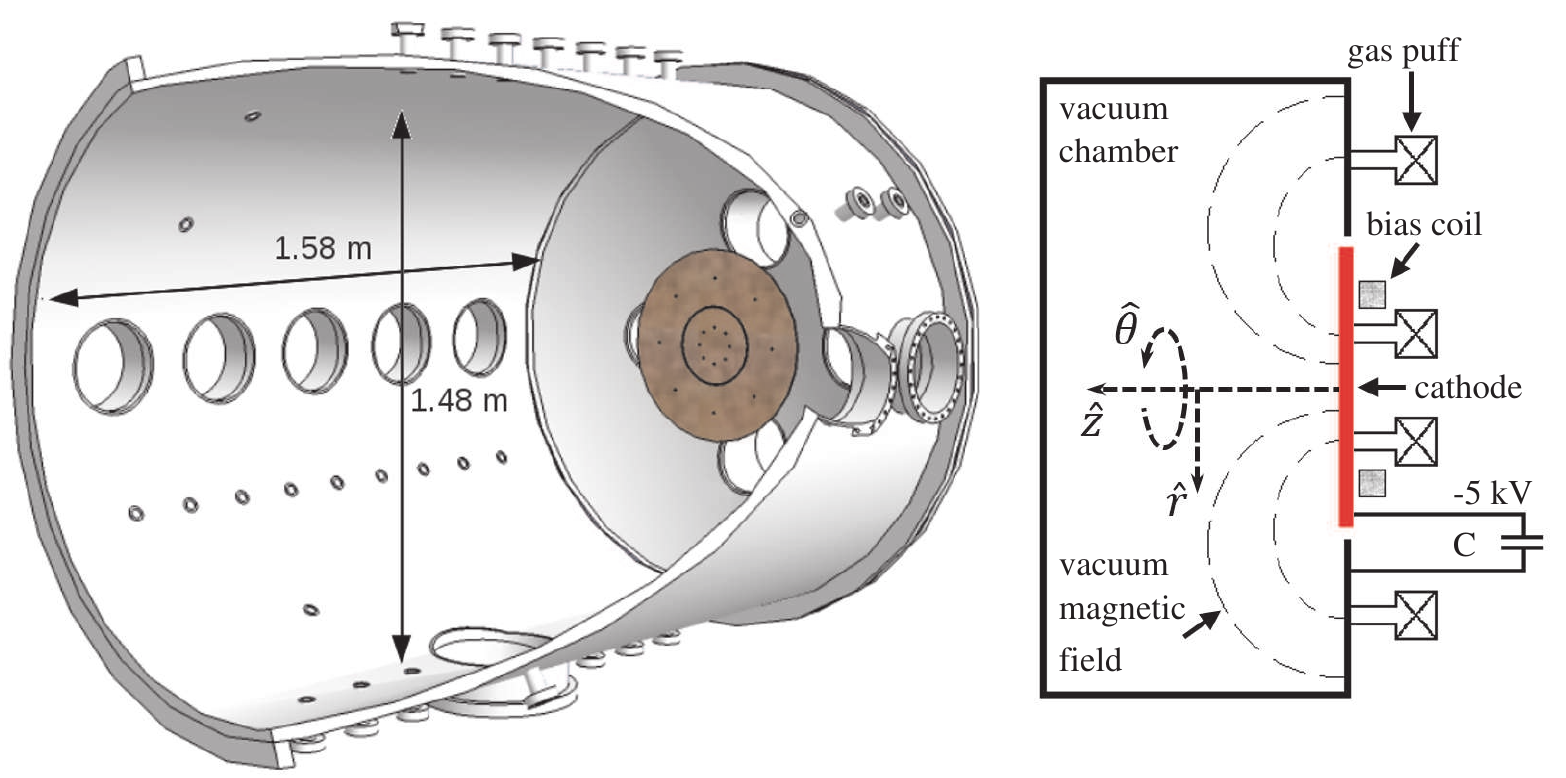} 
\plotone{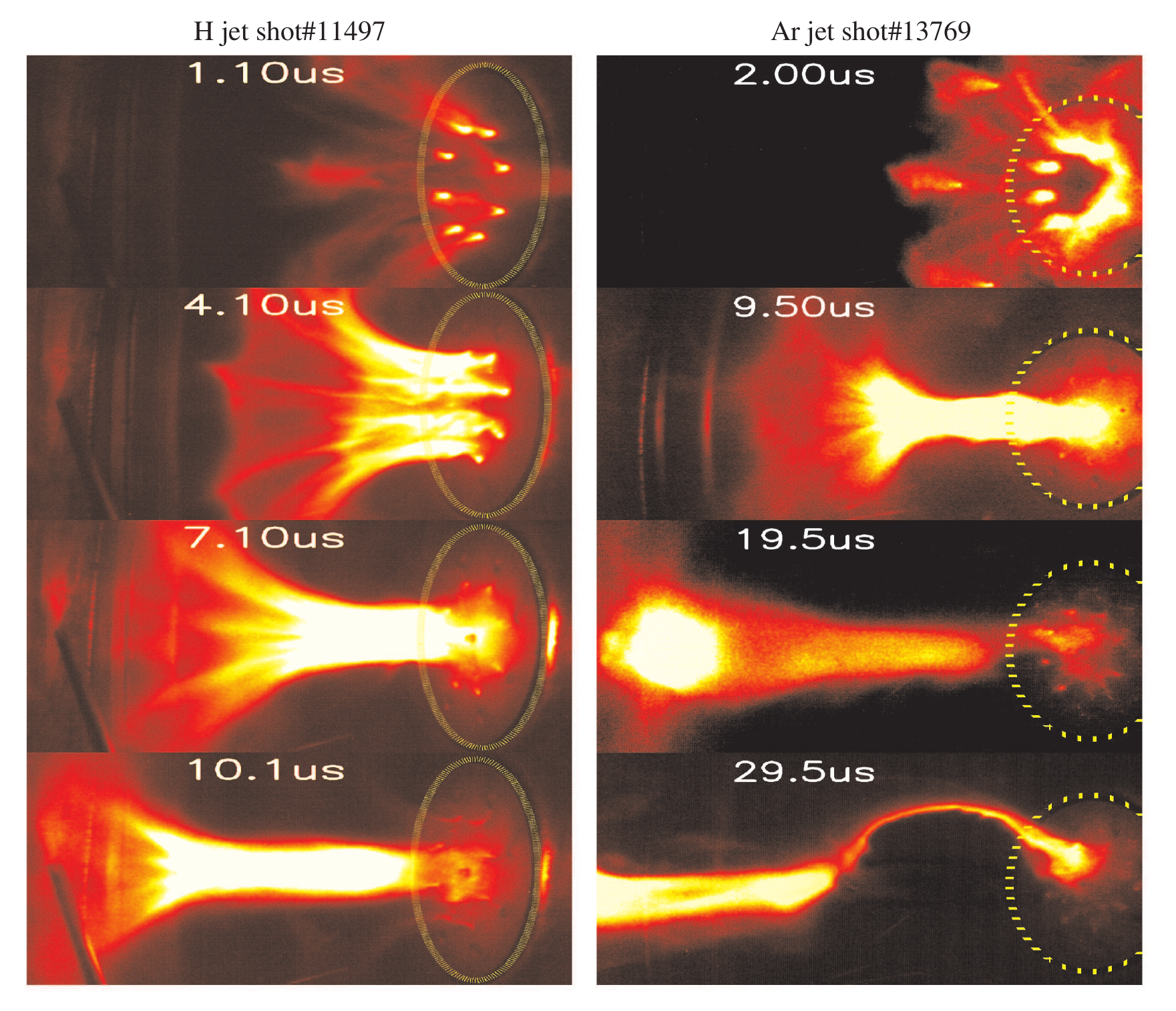} 
\caption{Upper left: 3D cross-sectional view of the vacuum chamber and the 
planar coaxial plasma gun. Upper right: sketch of the planar coaxial gun and 
the cylindrical coordinate. The central thick plane is the cathode. The sketch is not to scale. 
See Section 2 for detail. Lower panels: false color images showing the formation of 
a hydrogen plasma jet (left, shot 11497) and an argon plasma jet (right, shot 13769).
The hydrogen shot only used the $120$ $\mu$F $5$ kV power supply and the 
argon one used the PFN in addition to the power supply. The images are taken by a 
high-speed visible-light IMACON 200 camera at two slightly different angels. 
The dotted circles at the right of each frame is the $10$ cm radius 
central cathode.}\label{fig:expt}
\end{figure}

The Caltech experimental plasma jet is generated using a planar 
magnetized coaxial plasma gun mounted at one end of a $1.48$ m diameter, 
$1.58$ m long cylindrical vacuum chamber (sketch in Fig. \ref{fig:expt}). 
The vacuum pressure is $\sim 10^{-7}$ torr, corresponding to a background 
particle density of $3\times 10^{15}$ m$^{-3}$. 
The plasma gun has a $19.1$ cm diameter disk-shaped cathode and 
a co-planar annulus-shaped anode with 
inner diameter $d=20.3$ cm and outer diameter $D=51$ cm. The electrode plane is 
defined as $z=0$ and the central axis is the $z$ axis. At time $t=-10$ ms, 
a circular solenoid coil behind the cathode electrode generates a dipole-like 
poloidal background magnetic field for $\sim 20$ ms, referred to as the bias field. 
The total poloidal field flux is about $1.5$ mWb. At $t=-1$ to $-5$ ms, 
neutral gas is puffed into the vacuum chamber through eight evenly 
spaced holes at $r=5$ cm on the cathode and eight holes at $r=18$ cm on the 
anode at the same azimuthal angles. At $t=0$, a $120$ $\mu$F $5$ kV high 
voltage capacitor is switched across the electrodes. This breaks down the 
neutral gas into eight arched plasma loops spanning from the anode to the 
cathode following the bias poloidal field lines. At $0.6$ $\mu$s after breakdown, 
a $4$ kV pulse forming network (PFN) supplies additional energy to the plasma 
and maintains a total poloidal current at $60-80$ kA for $\approx 40$ $\mu$s. 
A typical current and voltage measurement is given in Fig. \ref{fig:VI_compare}.

Diagnostic instrumentation includes a high-speed visible-light IMACON 200 
camera, a 12-channel spectroscopic system \citep{Yun_Bellan_2010}, 
a He-Ne interferometer perpendicular to the jet \citep{Kumar_Bellan_2006}, 
a 20-channel 3D magnetic field probe array (MPA) along the $r$ direction with 
adjustable $z$ and $\sim 1$ $\mu$s response time \citep{Romero_MPA}, 
another similar MPA along the $z$ axis, a fast ion gauge, a Rogowski 
coil and a Tektronix high-voltage probe.

Fast ion gauge measurements show that the neutral particle number density 
immediately before the plasma breakdown is $10^{19}-10^{20}$ m$^{-3}$ \citep{AMoser_Thesis, Moser_Bellan_2012}. 
When the eight arched plasma loops are initially formed, the poloidal current and 
poloidal magnetic field in the loops are parallel to each other. However, the plasma is 
not a force-free system because of the toroidal magnetic field associated with 
the poloidal current. The inner segments of the eight arched loops, carrying parallel 
current from the anode to the cathode, mutually attract each other by the Lorentz 
force and merge into a single collimated plasma tube along the $z$ axis. 
A ten-fold density amplification in the jet due to collimation is 
observed by Stark broadening and interferometer measurements; 
these show the typical density of the collimated jet is $10^{22}-10^{23}$ m$^{-3}$ 
\citep{Yun_You_Bellan_2007, Kumar_Bellan_2009, Yun_Bellan_2010}. 
The poloidal magnetic field strength
in the plasma is also amplified from $< 0.05$ T to $\sim 0.2$ T, indicating that 
the field is frozen into the plasma and is collimated together with the plasma. 
This amplification of the magnetic field strength has also been observed 
spectroscopically \citep{Shikama_Bellan_2013}. The thermal pressure 
and axial magnetic field pressure $B_z^2/(2\mu_0)$
increase until they balance the radial Lorentz force and lead to a 
nearly constant jet radius of $2 -5$ cm (Fig. \ref{fig:expt}) and a 
toroidal magnetic field $B_{\theta}\sim 0.1-0.5$ T (see experimental 
measurements in Fig. \ref{fig:magnetic_structure}). This radial 
equilibrium is gradually established from small to large $z$, 
resulting in an MHD pumping mechanism that accelerates the plasma 
towards the $+z$ direction to form a jet. The typical jet velocity is 
$10-20$ km s$^{-1}$ for argon, $30-40$ km s$^{-1}$ for nitrogen and 
$\sim 50$ km s$^{-1}$ for hydrogen \citep{Kumar_Bellan_2009}. 
The plasma jet, as a one-end-free current-conducting plasma tube, 
undergoes a kink instability when its length grows long enough to 
satisfy the classical Kruskal-Shafranov kink threshold 
\citep{Hsu_Bellan_2003,Hsu_Bellan_2005}. When the kink grows 
exponentially fast and accelerates the plasma laterally away from the 
central axis, an effective gravitational force is experienced by the 
accelerating plasma jet. At the inner boundary of the kinked jet, 
where this effective gravity points from the displaced jet (dense plasma) to 
the $z$ axis (zero-density vacuum), a Rayleigh-Taylor instability 
occurs \citep{Moser_Nature}. The Rayleigh-Taylor instability eventually 
leads to a fast magnetic reconnection and destroys the jet structure. 
The jet life-time is $\sim 10$ $\mu$s for hydrogen, $20-30$ $\mu$s for 
nitrogen and $30-40$ $\mu$s for argon. Because heating is not important 
during this short, transient lifetime, the plasma remains at a relatively low 
temperature $T_e\sim 2$ eV inferred from spectroscopic 
measurements \citep{Yun_Bellan_2010}. Under typical experiment plasma
conditions, the temperature relaxation time between electrons and ions is about
$100$ ns, less than $1\%$ of the jet life time. Therefore $T_i\approx T_e\sim 2$ eV.
At this temperature, the plasma is essentially $100\%$ singly ionized
according to the Saha-Boltzmann theory, which is also confirmed by spectroscopic
measurements \citep[e.g., ][]{Yun_Bellan_2010, Hsu_Bellan_2003}.
Figure \ref{fig:expt} shows how the plasma is initially generated as 
eight arched loops, which then merge into one collimated jet. The jet then 
undergoes a kink instability when its length exceeds $\sim 30-40$ cm. 
For the current experiment configuration, the radius-length ratio of the jet 
in the final stage is about $1:10$.

For a typical experimental plasma with $n_e=10^{22}$ m$^{-3}$, 
$T_e=T_i=2$ eV, $B=0.2$ T and ion mass $\mu \equiv m_i/m_H$, 
the Debye length $\lambda_D\approx 10^{-7}$ m, the ion gyroradius 
$r_i\approx 0.7\sqrt{\mu}$ mm and the ion skin depth $d_i\approx 2\sqrt{\mu}$ 
mm are all significantly smaller than the length/radius of the experimental jet. 
The typical thermal to magnetic energy density ratio is $\beta\approx 0.1-1$, 
showing that the magnetic field is essential to the jet dynamics. 
Despite its relatively low temperature, the plasma has sufficiently high 
conductivity so that the Lundquist and magnetic Reynolds numbers $S\sim R_m\gtrsim 10^2\times(L/0.3\text{ m})/\sqrt{\mu}$ are both much greater than one with $L\sim 0.3$ m, where $L$ is the length scale of phenomena of interest. Therefore ideal MHD theory can describe jet 
global dynamics, such as collimation, acceleration and kinking 
\citep{Hsu_Bellan_2003, Hsu_Bellan_2005, Yun_You_Bellan_2007, 
Yun_Bellan_2010, Kumar_Bellan_2009, Kumar_Thesis}, and 
magnetic field diffusion is negligible during the jet dynamics. 
The kinked jet image in Fig. \ref{fig:expt} shows that the magnetic field is frozen into the plasma, consistent with ideal MHD theory. Hence the collimation 
of the bright plasma shown in Fig. \ref{fig:expt} also demonstrates the 
collimation of the magnetic field. The arched loops merging and the 
secondary Rayleigh-Taylor instability, on the other hand, involve ion skin depth length scale phenomena, that are smaller than can be described by MHD theory \citep{Moser_Nature}.

\section{Numerical MHD Simulations}
Discussion in this paper is restricted to the global 
axisymmetric behaviors of the jet, such as collimation and acceleration. 
Non-axisymmetric instabilities will be discussed in future publications. In this section, we prescribe appropriate initial and boundary conditions 
used to solve the ideal MHD equations numerically for the Caltech plasma jet 
experiment. 

\subsection{Normalization and Equations}

Number density, length and velocity are scaled to nominal reference values. 
In particular, density is normalized to $n_0=10^{19}$ m$^{-3}$, lengths are 
normalized to $R_0=0.18$ m (radial position of the outer gas feeding holes of the plasma gun in the experiment), 
and velocities are normalized to the ion sound speed $C_{s0}=\sqrt{2kT/m_i}=1.96\times10^4\sqrt{m_H/m_i}$ m s$^{-1}$ (with temperature $2$ eV). 
All other quantities are normalized to reference values derived from these 
three nominal values and ion mass $m_i$. Table \ref{tab:units} lists the 
derivation and the normalization values adopted in the experimental 
hydrogen/argon jet simulation and the AGN jet simulation by 
\citet{Li2006_MagneticTower}. SI units are used for the lab experiment 
while cgs units are used for the AGN jet in order to facilitate comparison to respective experimental and astrophysical literature.


\begin{deluxetable}{lcccc}
\centering
\tablecaption{Normalization units for Experimental H/Ar Jet Simulation and AGN Jet Simulation}
\tablenum{1}
\tablehead{\colhead{Quantity unit} & \colhead{Quantity symbols} & \colhead{H ($\mu=1$)} & \colhead{Ar  ($\mu=40$)} & \colhead{AGN jet ($\mu=1$)}}
\startdata
Length   		& $R_0$ 	& $0.18$ m 						& $0.18$ m						& $15$ kpc\\
Number density	& $n_0$ 	& $10^{19}$ m$^{-3}$			& $10^{19}$ m$^{-3}$			& $3\times10^{-3}$ cm$^{-3}$\\
Speed			& $C_{s0}$	& $1.96\times 10^{4}$ m s$^{-1}$ 	& $3.1\times10^3$ m s$^{-1}$			& $1.16\times10^8$ cm s$^{-1}$\\
Ion weight 	& $\mu=m_i/m_H$ 	& $1$				& $40$	& $1$\\
\hline
Time 			& $t_0=R_0/C_{s0}$ 	& $9.2$ $\mu$s 					& $58.2$ $\mu$s					& $1.3\times10^{7}$ yr\\
Mass density 	& $\rho_0=n_0m_i/2$  & $8.4\times 10^{-9}$ kg m$^{-3}$	& $3.3\times10^{-7}$ kg m$^{-3}$	& $2.5\times10^{-30}$ g cm$^{-3}$\\
Pressure 		& $p_0=\rho_0C_{s0}^2$ 	& $3.2$ pa						& $3.2$ pa						& $3.4\times10^{-11}$ erg cm$^{-3}$\\
Temperature		& $k_BT=m_iC_{s0}^2/2$	& $2$eV							& $2$eV							& $7$ keV\\
Energy  		& $E_0=p_0R_0^3$ 	& $0.0187$ J 					& $0.0187$ J					& $3.4\times10^{57}$ erg\\
Power  			& $P_0=E_0/t_0$ 	& $2.0\times10^3$ Watt			& $321$ Watt					& $2.6\times10^{50}$ erg/yr\\
Magnetic field  & $B_{0}=\sqrt{\mu_0p_0}$ 	& $0.002$ T						& $0.002$ T						& $2\times10^{-5}$ Gauss\\
Magnetic flux   & $\Psi_{0}=B_0R_0^2$& $0.0648$ mWB					& $0.0648$ mWB					& $4.4\times10^{40}$ G cm$^2$\\
Current density & $J_0=B_0/(\mu_0R_0)$ 	& $8.871\times10^3$ A m$^{-2}$		& $8.871\times10^3$ A m$^{-2}$		& $3.5\times10^{-28}$ A cm$^{-2}$\\
Current 		& $I_0=J_0R_0^2$ 	& $2.874\times10^2$ A			& $2.874\times10^2$ A			& $7.6\times10^{17}$ A\\
Voltage 		& $V_0=P_0/I_0$ 	& $7.07$ V						& $1.118$ V						& $1.1\times10^{18}$ V\\
\enddata
\label{tab:units}
\end{deluxetable}

\placetable{tab:units}

The dimensionless ideal MHD equations, normalized to the quantities given in Table \ref{tab:units}, are
\begin{mathletters}
\begin{eqnarray}
\frac{\partial\rho}{\partial t} +\nabla\cdot (\rho\mathbf{v}) =0& \label{eqn:continuity}\\
\frac{\partial (\rho\mathbf{v})}{\partial t}+\nabla\cdot(\rho\mathbf{v}\mathbf{v}+P_g\overleftrightarrow{\mathbf{I}}+P_B\overleftrightarrow{\mathbf{I}}-\mathbf{B}\mathbf{B})=0& \label{eqn:momentum}\\
\frac{\partial e}{\partial t}+\nabla\cdot[(e+P_g+P_B)\mathbf{v}-\mathbf{B}(\mathbf{v}\cdot\mathbf{B})]=\dot{e}_{\rm inj}& \label{eqn:energy}\\
\frac{\partial\mathbf{B}}{\partial t}-\nabla\times(\mathbf{v}\times\mathbf{B})=\dot{\mathbf{B}}_{\rm inj}&\label{eqn:induction}
\end{eqnarray}
\end{mathletters}
\noindent where all the dimensionless variables have their conventional meaning. 
The momentum equation and the energy equation have been written in the form of conservation laws. 
We assume the same ion/electron temperature $T=T_i=T_e$. The particle number density 
$n=2n_e=2n_i$ is used assuming singly-ionized plasma. The ionization status is assumed to be time-independent. The equation of state for an ideal gas 
with adiabatic index $\gamma=5/3$ is used. The gas pressure $P_g=n_ik_BT_i+n_ek_BT_e=nk_BT$ 
is then related to the thermal energy density by $e_{\rm thermal}=P_g/(\gamma-1)$. The magnetic 
pressure $P_B$, or magnetic energy density $e_B$, is $P_B=e_B=B^2/(2\mu_0)$ and 
the total energy density is $e\equiv\rho v^2/2+P_g/(\gamma-1)+P_B$.

An injection term $\dot{\mathbf{B}}_{\rm inj}$ is added to the induction equation. The associated dimensionless energy density injection is
\begin{equation}
\dot{e}_{\rm inj}=\dot{\mathbf{B}}_{\rm inj}\cdot\mathbf{B},\label{eqn:E_inj}
\end{equation}
where $\mathbf{B}$ is the magnetic field.

Simulations are performed in a 3D Cartesian coordinate system 
$\{x,y,z\}$ 
using the 3D MHD code as part of the Los Alamos
COMPutational Astrophysics Simulation Suite \citep[LA-COMPASS, ][]{LiLi2003_AMR3D}. 
This package was previously used for simulating AGN jets \citep[e.g.,][]{Li2006_MagneticTower}. 
The solving domain is a cube $[-4R_0,4R_0]^3=[-0.72$ m$,0.72$ m$]^3$, 
similar to the vacuum chamber size in the experiment. Each Cartesian axis 
is discretized into $800$ uniformly spaced grids, giving a total of $5.12\times 10^8$ grid points. 
The spatial resolution $\Delta x=8R_0/800=1.8$ mm in the simulation is significantly greater 
than the Debye length, and is similar to the ion gyroradius and the ion skin depth of the plasma 
jet in the experiment. A typical run takes 5 to 24 hr on the Los Alamos National Lab 
Turquoise Network using 512 processors.

In contrast to the experiment where the jet exists only for positive $z$, the simulation 
has a mirrored plasma jet in the negative $z$ direction so as to have a bipolar 
system centered at $z=0$ plane. The solving domain contains plasma 
only and has no plasma-electrode interaction region. Non-reflecting outflow 
boundary conditions are imposed at the boundaries (large $x$, $y$ or $z$). The 
MHD equations are solved in Cartesian coordinates so that no computational 
singularity exists at the origin.

\subsection{Initial Condition}
\subsubsection{Initial Global Poloidal Magnetic Field}
It is generally believed that the poloidal and toroidal magnetic component evolve together under the dynamo processes in accretion disk and surrounding corona. However, when the poloidal component varies slower than the toroidal component, it is possible to treat the two components separately. In \citet{Lynden-Bell_1996, Lynden-Bell_2003}, a poloidal field is assumed to be pre-existing, and the toroidal field is generated by twisting the upward flux relative to downward flux. During this process, the poloidal flux remains constant while toroidal field is enhanced with the increase of number of turns (helicity). These processes are realized equivalently in the lab experiment, where an initial dipole poloidal field is first generated by an external coil, and then helicity is increased by injecting poloidal current. In the simulation, an initial dipole poloidal magnetic field is similarly imposed, given by
\begin{equation}
\Psi_{\rm pol}(r,z)\equiv 2\pi\alpha_p\frac{r^2}{(l^2+a_0^2)^{3/2}} e^{-l^2},\label{eqn:init_PolFlux}
\end{equation}
where $a_0\equiv 0.623R_0=11.2$ cm ($R_0=0.18$ m, see Table \ref{tab:units}) and 
$l\equiv\sqrt{r^2+z^2}$ is the distance from the origin. This configuration is topologically 
similar to the initial poloidal flux $\Psi_{\rm pol}=r^2e^{-l^2}$ adopted by 
\citet{Li2006_MagneticTower}.
By default, simulation equations/variables will be written in dimensionless form with reference units given in Table \ref{tab:units}. For example, Eq. \ref{eqn:init_PolFlux} is the dimensionless version of $\Psi_{\rm pol}(r,z)=2\pi\alpha_p B_0R_0^2 (r/R_0)^2/[(l/R_0)^2+(a_0/R_0)^2]^{3/2}e^{-l^2/R_0^2}$, where $B_0$ and $R_0$ are given in Table \ref{tab:units}. Compared to the ideal infinitesimal magnetic dipole flux $\Psi\propto r^2/l^3$, 
$\Psi_{\rm pol}$ contains a constant factor $a_0$ to make the dipole source finite; 
it also has an exponential decay at large distance so that the initial field vanishes 
at the solving domain boundaries. At small $r$ and $z$, $\Psi_{\rm pol}\propto r^2$ 
hence $B_z$ is nearly constant. $a_0$ is selected so that 
$\Psi_{\rm pol}(r=r_1,z=0)=\Psi_{\rm pol}(r=r_2,z=0)$, where $r_1=0.278\Rightarrow5$ cm 
and $r_2=1\Rightarrow18$ cm corresponding to the radii of the inner and 
outer gas lines in the experiment. The dimensionless parameter $\alpha_p$ 
quantifies the strength of the flux.
The vector potential can be selected to be $\mathbf{A}=(\Psi_{\rm pol}/(2\pi r))\hat{\theta}$. 
The initial poloidal field is
\begin{gather}
\mathbf{B}_{\rm pol}=\nabla\times\mathbf{A}=\frac{1}{2\pi}\nabla \Psi_{\rm pol} \times \nabla\theta\label{eqn:B_pol}\\
\Rightarrow
\left\{
\begin{split}
B_{r}&=\frac{\alpha_pzre^{-l^2}}{(l^2+a_0^2)^{5/2}}(3+2a_0^2+2l^2)\\
B_{z}&=\frac{\alpha_pe^{-l^2}}{(l^2+a_0^2)^{5/2}}\left[2(1-r^2)(l^2+a_0^2)-3r^2 \right]
\end{split}
\right.
\label{eqn:B_pol_comp}
\end{gather}
where $\hat{\theta}$ is the azimuthal unit vector and $\nabla\theta=\hat{\theta}/r$. The total poloidal flux is
\begin{equation}\label{eqn:tot_flux}
\Psi_{\rm  0,pol}\equiv\Psi_{\rm pol}(r_{o},0)=2.448\alpha_p\Rightarrow 0.1586\alpha_p\text{ mWb},
\end{equation}
where $r_{o}=0.5667\Rightarrow10.20$ cm is the position of the null of the initial 
poloidal field, i.e., $B_{z}(r_o,0)=0$. The first frame of Fig. \ref{fig:den_evolution}
shows the flux contours of the initial poloidal field in the $rz$ plane.

The toroidal current associated with the poloidal field is
\begin{equation}\label{eqn:J_tor}
J_{\theta}=\partial_z B_r-\partial_r B_z=-\alpha_p\frac{re^{-l^2}}{(l^2+a_0^2)^{7/2}}\cdot g(l)
\end{equation}
where
\begin{equation}
g(l)=4l^6+(8a_0^2+2)l^4+4a_0^2(a_0^2-2)l^2-5a_0^2(a_0^2+3).
\end{equation}
Simple calculation shows that $l_0=0.9993\approx 1$ is the only zero point of 
$g(l)$ and $g(l)<0$ for $0\le l<l_0$ and $g(l)>0$ for $l>l_0$.

\subsubsection{Initial Mass Distribution}

In the experiment, plasma is initially created following the path of initial poloidal field lines (see Fig. \ref{fig:expt} H jet at $1.1$ $\mu$s and Ar jet at $2.0$ $\mu$s), i.e., the plasma is distributed around the $\Psi_{\rm pol}(r,z)=\Psi_0$ surface. Here $\Psi_{\rm pol}(r,z)$ is the initial poloidal flux function (Eqn. \ref{eqn:init_PolFlux}) and $\Psi_0\equiv \Psi_{\rm pol}(r_1,0)=\Psi_{\rm pol}(r_2,0)$ is the flux contour connecting the inner ($r_1=5$ cm) and outer ($r_2=18$ cm) gas feeding holes. A possible choice for the initial mass distribution function in the simulation is $n_{\rm init}\sim \exp[-\delta(\Psi_{\rm pol}(r,z)-\Psi_0)^2]$.

Note that this initial distribution has low plasma density on the axis. In the experiment, fast magnetic reconnection occurs as the eight arched loops merge into one. This allows the plasma and magnetic field to fill in the central region. The ideal MHD simulation, however, lacks the capability to simulate the fast magnetic reconnection, and hence cannot accurately describe the merging process. 
As a compromise, we start the simulation immediately after the merging process but before the collimation and propagation processes.
We therefore choose a simple form topologically similar to the contour $\Psi_{\rm pol}(r,z)=\Psi_0$ but without the central hollow region, namely
\begin{equation}
n_{\rm init}(r,z)=1+n_{\rm init,0}\cdot e^{-l^2}\cdot e^{-\delta\left[(r-1/2)^2+z^2-1/4\right]^2}.\label{eqn:premerged_mass}
\end{equation}
The first term $1$ corresponds to a background particle number density 
$10^{19}$ m$^{-3}$. This is $\sim 10^3$ times more dense than the 
background in the experiment, but still $\sim 10^{-3}$ less dense than 
the plasma jet. $n_{\rm init,0}$ is the assumed initial plasma number density. 
The $e^{-\delta\left[(r-1/2)^2+z^2-1/4\right]^2}$ term states that the plasma 
is initially distributed over a torus surface $(r-1/2)^2+z^2=1/4$ connecting 
$r=0$ and $r=1=18$ cm at mid plane. The torus surface is roughly parallel to the initial 
poloidal flux surface $\Psi_{\rm pol}(r,z)=\Psi_0$, but without the central hole. 
The $e^{-l^2}$ term assures that the initial plasma 
is localized around the origin. Using this distribution, the central 
region $r\simeq 0$ in the simulation is initially filled with dense plasma.

\subsection{Helicity and Energy Injection}
\subsubsection{Compact Injection Near the $z=0$ Plane}
Toroidal magnetic flux is continuously injected into the simulation system, 
in order to replicate the energy and magnetic injection through the electrodes 
in the experiment. The helicity conservation equation in an ideal MHD plasma 
with volume $\mathcal{V}$ is 
\begin{equation}
\frac{dK_{rel}}{dt}=-\int_{\partial \mathcal{V}}(2V\mathbf{B})\cdot d\mathbf{S}=2\Psi_{\rm pol}\cdot\frac{\partial (IL)}{\partial t},\label{eqn:helicity_injection}
\end{equation}
where $K_{\rm rel}$ is the relative magnetic helicity, $\partial \mathcal{V}$ is the 
boundary of the volume and the area $d\mathbf{S}$ is normal to the boundary, 
$V$ is the electrode voltage, $I$ is the total current through the plasma and $L$ is 
the plasma self inductance across the electrodes \citep{Finn_Antonsen_1985,Berger_1999,Bellan_Spheromaks,Kumar_Thesis}. 
The electrode surface in the experiment is the effective $\partial \mathcal{V}$. 
When a poloidal magnetic field is present, Eq. (\ref{eqn:helicity_injection}) states 
that magnetic helicity injection can be realized either by maintaining a non-zero 
voltage across the electrodes, or by increasing the poloidal current/toroidal field in the plasma. 
In the experiment, these two methods are essentially equivalent. Meanwhile, 
magnetic energy is also injected into the plasma by 
$\dot{E}=P=IV=Id\Psi_{tor}/dt$, 
where $\Psi_{tor}$ is the toroidal magnetic flux. Since neither electric field nor potential 
is explicitly used in the simulation, we choose the second method to inject helicity. 
Thus we inject toroidal magnetic field into the system to increase the poloidal current 
and the magnetic helicity. 
The toroidal field injection term in Eq. (\ref{eqn:induction}) is defined as
\begin{equation}
\dot{\mathbf{B}}_{\rm inj}\equiv\gamma_b(t)\mathbf{B}_{\rm tor}\label{eqn:Binj_define},
\end{equation}
where $\gamma_b(t)$ is the injection rate and
\begin{equation}
\mathbf{B}_{\rm tor}=\frac{f(\Psi_{\rm pol})}{2\pi r}e^{-Az^2}\hat{\theta}=\frac{1}{2\pi}f(\Psi_{\rm pol})e^{-Az^2}\nabla\theta
\end{equation}
is a pure toroidal field. The localization factor $A$ is a large positive number 
so that toroidal field injection is localized near the $z=0$ plane. $f(\Psi_{\rm pol})$ is 
an analytical function of $\Psi_{\rm pol}$ and following the magnetic tower model 
used in \citet{Li2006_MagneticTower}, we choose $f(\Psi_{\rm pol})=\alpha_t\Psi_{\rm pol}$ 
so that
\begin{equation}
\mathbf{B}_{\rm tor}=\alpha_t\alpha_p\frac{r}{(l^2+a_0^2)^{3/2}}e^{-l^2}e^{-Az^2}=\alpha_t\alpha_p\frac{r}{(r^2+z^2+a_0^2)^{3/2}}e^{-r^2-(A+1)z^2}.\label{eqn:B_tor}
\end{equation}
The poloidal current associated with this toroidal field is
\begin{equation}
\mathbf{J}_{\rm pol}=\nabla\times\mathbf{B}_{\rm tor}=\frac{1}{2\pi}\nabla\left( \alpha_t\Psi_{\rm pol}e^{-Az^2}\right)\times\nabla\theta=\alpha_te^{-Az^2}\mathbf{B}_{\rm pol}+\frac{\alpha_t\Psi_{\rm pol}}{\pi r}Aze^{-Az^2}\hat{r},\label{eqn:J_pol}
\end{equation}
where $\nabla\times\nabla\theta=0$ and Eq. (\ref{eqn:B_pol}) are used.

At $z=0$, $B_{\rm tor}=\alpha_t\Psi_{\rm pol}/(2\pi r)$. Therefore the net poloidal current within radius $r$ is $2\pi rB_{\rm tor}=\alpha_t\Psi_{\rm pol}$. Using Eqn. \ref{eqn:tot_flux}, the total positive poloidal current associated with $B_{\rm tor}$ is
\begin{equation}\label{eqn:pol_current}
I_{\rm pol}=2.488\alpha_t\alpha_p\Rightarrow 0.704\alpha_t\alpha_p \text{ kA}
\end{equation}
The localization factor $A$ has no impact on the total poloidal current.

It is important to point out that the field injection term in the induction Eq. (\ref{eqn:induction}) 
is a compromise used to avoid having a plasma-electrode interaction boundary condition. 
Theoretically, Eq. (\ref{eqn:induction}) is not physically correct because of the injection term. 
However, because the localization factor $A$ is a large positive number, the magnetic 
energy of $\mathbf{B}_{\rm tor}$ decreases rapidly with $z$. Therefore the ``unphysical" 
region is very localized to the vicinity of the $z=0$ plane. In particular, using $A=9$, 
the total toroidal magnetic energy at the $z=0.307\Rightarrow5.5$ cm plane is 
only $10\%$ of the total planar magnetic energy at the $z=0$ plane. 
The toroidal magnetic flux within $|z|<0.307$ contributes $87\%$ of the total toroidal 
flux, although the volume is only $7.7\%$ of the total simulation domain. 
We define $z_{\rm foot}\equiv 0.307$, so the region where $|z|<z_{\rm foot}$ is the ``engine region" 
where most of the energy injection is enclosed, and the region outside the 
engine region ($|z|>z_{\rm foot}$) is the ``jet region" where unphysical toroidal field 
injection does not occur. In the engine region, the toroidal magnetic field is directly 
added to the existing configuration by the modified induction equation 
(Eq. \ref{eqn:induction}). The injection also adds magnetic helicity, poloidal current and 
magnetic energy. In the jet region, on the other hand, this direct injection is negligible 
so the ideal MHD laws hold almost perfectly. The helicity, current and energy enter 
the jet region with the plasma flow.

In the simulations presented here, we use $A=9$. Although the choice of $A$ is 
somewhat arbitrary, in general, $A$ needs to be sufficient large to localize the 
engine region to the vicinity of the $z=0$ plane. This compact engine region serves 
as an effective plasma-electrode interface, and leaves most of the simulation 
domain described by the correct induction equations (i.e., no artificial injection). 
If a small $A$ were used, injection would occur globally. There would then be a 
large amount of energy directly added to remote regions with low density plasma. 
A magnetized shock would then arise and dissipate injected energy. Using a large 
$A$ guarantees that the magnetic field is mostly frozen into the dense plasma. 
However, $A$ should be not too large in our simulation, since otherwise numerical 
instability and error would occur because of excessive gradients.

The process of helicity/energy injection in the simulation is not exactly the 
same as in the experiments or the astrophysical case. In the experiment, a 
non-zero electric potential drop between the electrodes is responsible for 
the process. In AGN jet or stellar jet cases, the injection process could also be
accompanied by electric potential drop in the radial direction as a result of 
interaction among the central object, wind, magnetic field and the accretion disk 
dynamics, such as differential rotation of the disk and corona. However, the artificial injection of a purely toroidal field should produce mathematically equivalent magnetic structure. This injection is also consistent with the asymptotic X-winds solution by \citet{Shu_1995} and \citet{Shang_2006}.

\subsubsection{Jet Collimation as a Result of Helicity/Energy Injection}

To illustrate how injected toroidal magnetic field impacts the system, we consider a
 ``virtual magnetic field" configuration composed by $\mathbf{B}_{\rm pol}$ 
 (defined in Eq. \ref{eqn:init_PolFlux}-\ref{eqn:B_pol_comp}) and 
 $\mathbf{B}_{\rm tor}$ (defined in Eq. \ref{eqn:B_tor}).
The Lorentz force
\begin{equation}
\begin{split}
&\mathbf{F}\equiv \mathbf{J}\times\mathbf{B}=(\mathbf{J}_{\rm pol}+\mathbf{J}_{\rm tor})\times(\mathbf{B}_{\rm pol}+\mathbf{B}_{\rm tor}) 
\\
&\mathbf{F}_{\rm pol}=\mathbf{J}_{\rm pol}\times\mathbf{B}_{\rm tor}+\mathbf{J}_{\rm tor}\times\mathbf{B}_{\rm pol}\\
&\mathbf{F}_{\rm tor}=\mathbf{J}_{\rm pol}\times\mathbf{B}_{\rm pol}\label{eqn:JxB}
\end{split}
\end{equation}
has both poloidal and toroidal components.

We first examine the toroidal component of the Lorentz force, or, the twist force. 
The first component of $\mathbf{J}_{\rm pol}$ in Eq. (\ref{eqn:J_pol}) is parallel to 
$\mathbf{B}_{\rm pol}$, hence only the second term contributes to the twist, namely,
\begin{equation}
\mathbf{F}_{\rm tor}=\frac{\alpha_t\Psi_{\rm pol}Aze^{-Az^2}}{\pi r}\hat{r}\times B_z\hat{z}
=-2\alpha_t\alpha_p^2A\frac{rz}{(l^2+a_0^2)^4}e^{-2l^2-Az^2}
\cdot\left[2(1-r^2)(l^2+a_0^2)-3r^2\right]\hat{\theta}.
\label{eqn:F_tor}
\end{equation}
For small radius, the twist force scales as $F_{\rm tor}/r\sim ze^{-(A+2)z^2}/(z^2+a_0^2)^3$. 
The twist force is strongest at $z=0.166\Rightarrow3$ cm and weak for very small $z$ and 
large $z$. In the simulation, $F_{\rm tor}$ twists the plasma differently at different radii and height, and hence contributes to $E_r$ by increasing $v_{\theta}B_z$ negatively. This electric field 
is equivalent to the voltage across the inner cathode and outer anode in the experiment.

In the poloidal component of the Lorentz force, the $\mathbf{J}_{\rm tor}\times\mathbf{B}_{\rm pol}$ 
term is the hoop force that expands the system resulting from the poloidal magnetic field; 
while $\mathbf{J}_{\rm pol}\times\mathbf{B}_{\rm tor}$ is the pinch force and is caused 
solely by the toroidal magnetic field.
Insertion of Eq. (\ref{eqn:B_pol},\ref{eqn:B_pol_comp},\ref{eqn:J_tor},\ref{eqn:J_pol},\ref{eqn:B_tor}) 
into Eq. (\ref{eqn:JxB}) yields

\begin{align}
F_r=&-\frac{\alpha_p^2re^{-2l^2}}{(l^2+a_0^2)^6}\left[2(1-r^2)(l^2+a_0^2)-3r^2\right]\cdot\left[g(l)+\alpha_t^2e^{-2Az^2}(l^2+a_0^2)^2\right]\label{eqn:Fr}\\
F_z=&\frac{\alpha_p^2r^2ze^{-2l^2}}{(l^2+a_0^2)^6}\left[\left(3+2l^2+2a_0^2\right)g(l)+\alpha_t^2e^{-2Az^2}(l^2+a_0^2)^2(3+4l^2+4a_0^2)\right].\label{eqn:Fz}
\end{align}
The terms containing $g(l)$ result from the poloidal field and the terms proportional to 
$\alpha_t^2$ are given by the toroidal field. In the small $\alpha_t$ limit, the pinch applied 
by the toroidal field is weak, so the $g(l)$ term determines the direction of the Lorentz force. 
In the region of small $r$ and $l$, $F_r>0$ and $F_z/z<0$, showing that the plasma expands 
and is made more diffuse by the hoop force. The same argument is true for $l<l_0\approx1$ 
and for finite $\alpha_t$ with $Az^2\gg1$. In the cases where $\alpha_t$ is sufficiently large, 
i.e., the pinch due to the poloidal current/toroidal field overcomes the hoop force, 
$F_r<0$ and $F_z/z>0$ for small $r$. This is where the toroidal field squeezes the plasma 
radially and lengthens it axially. To see this more clearly, if we ignore the poloidal field 
effect by dropping the terms containing $g(l)$, the radial Lorentz force is 
$F_r/r\propto e^{-2(A+1)z^2}/(l^2+a_0^2)^3$, which decreases rapidly along the 
$z$ axis. Hence the plasma is pinched and pressurized more at small $z$ than 
at large $z$. The huge pressure gradient along the central axis, due to the huge 
gradient of collimation force, then accelerates the plasma away from the $z=0$ plane. 
Equivalently, the large gradient of the toroidal magnetic pressure $B_{\theta}^2/(2\mu_0)$ 
in the $z$ direction is responsible for the collimation and acceleration of the plasma.
 
It is important to point out that the Lorentz force is primarily poloidal. Since
\begin{equation}
\frac{F_{\rm tor}}{F_r}=\frac{2\alpha_tAze^{-Az^2}(l^2+a_0^2)^2}{g(l)+\alpha_t^2e^{-2Az^2}(l^2+a_0^2)^2}\propto\left\{
\begin{aligned}
\alpha_t& \text{  for small } \alpha_t\\
1/\alpha_t& \text{  for large } \alpha_t
\end{aligned}
\right.,\label{eqn:Ftor_Fr_ratio}
\end{equation}
$F_r$ is usually much stronger than $F_{\rm tor}$.

The above analyses show the Lorentz force tends to squeeze the plasma 
radially and accelerate it axially with the presence of $\mathbf{B}_{\rm tor}$. 
However, in the simulation, only $\mathbf{B}_{\rm pol}$ is initially imposed as the 
bias poloidal field. The toroidal field is continuously injected into the system 
at small $z$. Meanwhile, the existing poloidal and toroidal magnetic field 
configuration is continuously deformed together with the plasma. 
Eq. (\ref{eqn:F_tor}-\ref{eqn:Fr}) are not exact expressions of the Lorentz 
force experienced by the plasma. However, Eq. (\ref{eqn:F_tor}-\ref{eqn:Fr}) 
nevertheless gives a semi-quantitative description of how injected 
toroidal field affects the plasma.

In summary, we have established both the initial condition and the continuous injection 
condition for simulating the Caltech plasma jet experiment. Only a poloidal field and a 
dense plasma distributed roughly parallel to the field lines are imposed initially. 
As the plasma starts to evolve, although the hoop force by the initial toroidal 
current tries to expand the plasma radially, the injected toroidal magnetic field 
(poloidal current) applies Lorentz force that overcomes the poloidal field pressure, 
and squeezes the plasma radially and lengthens it axially to form a jet in both 
the $+\hat{z}$ and $-\hat{z}$ directions. We only consider the $+z$ part as 
the $-z$ part is a mirror image.

\section{Simulation results}
In this section, we present some typical simulation results and compare them to the experimental results.

\subsection{A Typical Argon Jet Simulation}
First we show a typical argon plasma jet simulation ($\mu=40$). The initial poloidal 
flux factor is selected as $\alpha_p=10$ corresponding to a $1.59$ mWb poloidal flux 
with maximum $B_z$ strength of $0.165$ T at the origin. The initial mass distribution is 
given by Eq. (\ref{eqn:premerged_mass}) with $\delta=40$ and $n_{\rm init,0}=4000$, 
corresponding to a maximum initial electron number density $2\times10^{22}$ m$^{-3}$.

The dimensionless injection coefficient is
\begin{equation}
\gamma_b(t)\alpha_t=1000e^{-30t}+150e^{-20(t-0.5)^2}\label{eqn:injection_coef}
\end{equation}
for $0\le t\le 0.6=35$ $\mu$s, which contains a short exponential decay and 
then a long-duration Gaussian profile. This corresponds to the fast power input 
by the main capacitor and then the long-duration power input by the PFN in the experiment. This injection rate is obtained based on the experiment
current characteristics. In the experiment, the main capacitor gives rise to a plasma
poloidal current at a rate of $\approx150$ kA$/3$ $\mu$s 
$\times(\pi/2)\sim 10^2$ kA$/\mu$s. The PFN supplies current $60-80$ kA with a rise
time of $\sim 10$ $\mu$s, giving a current injection rate $\sim 10$ kA$/\mu$s. With $\alpha_p=10$,
Eqn. \ref{eqn:pol_current} indicates a dimensionless injection rate $\gamma_b\alpha_t$ $\sim 10^3$
for the main capacitor and $\sim 10^2$ for the PFN.

The localization factor is $A=9$ so the engine region extends up to $z_{\rm foot}=0.307\Rightarrow 5.5$ cm. 
The initial plasma temperature is uniformly $T_i=T_e=2$ eV, and the plasma remains $100\%$ singly ionized through the simulation.


\subsubsection{Global Energy Analysis}

\begin{figure}
\figurenum{2}
\epsscale{1}
\plotone{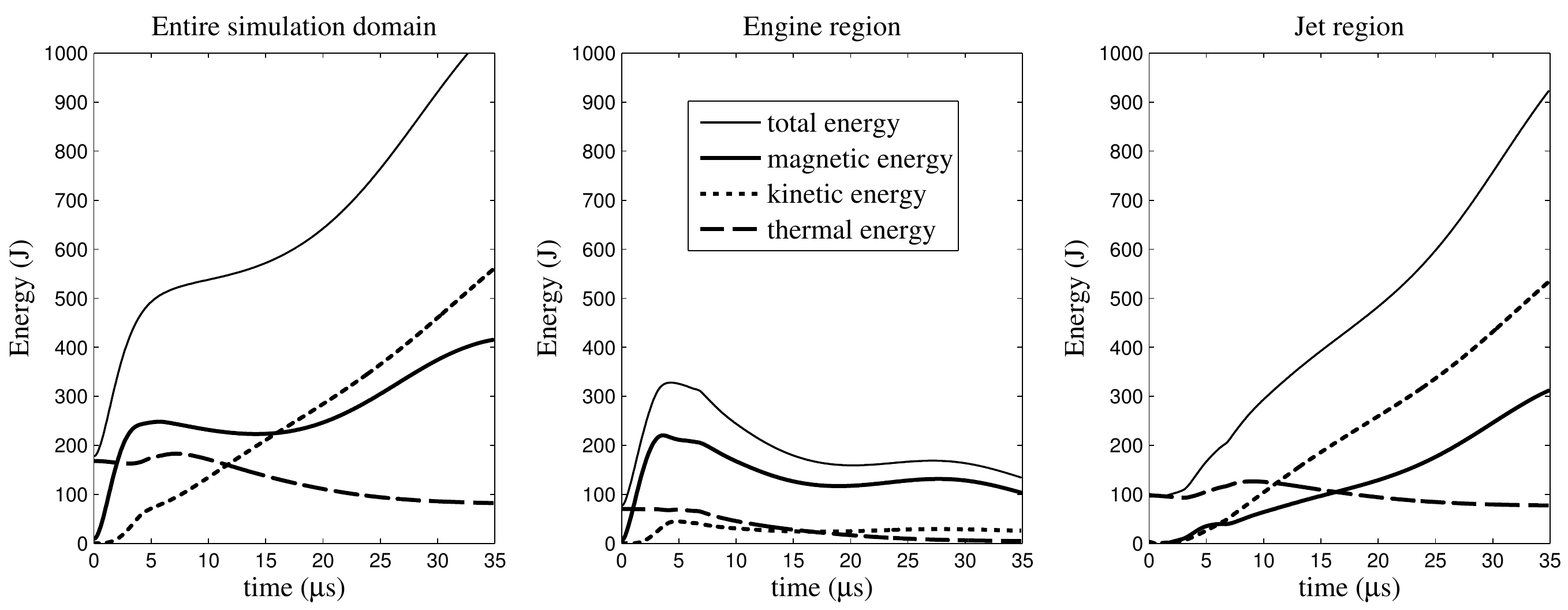} 
\caption{Evolution of different energy components in the entire simulation 
domain (left), engine region $|z|<z_{\rm foot}$ (middle) and  
jet region $|z|\geq z_{\rm foot}$ (right).}\label{fig:energy_partition}
\end{figure}

\begin{figure}[t!]
\figurenum{3}
\epsscale{0.8}
\plotone{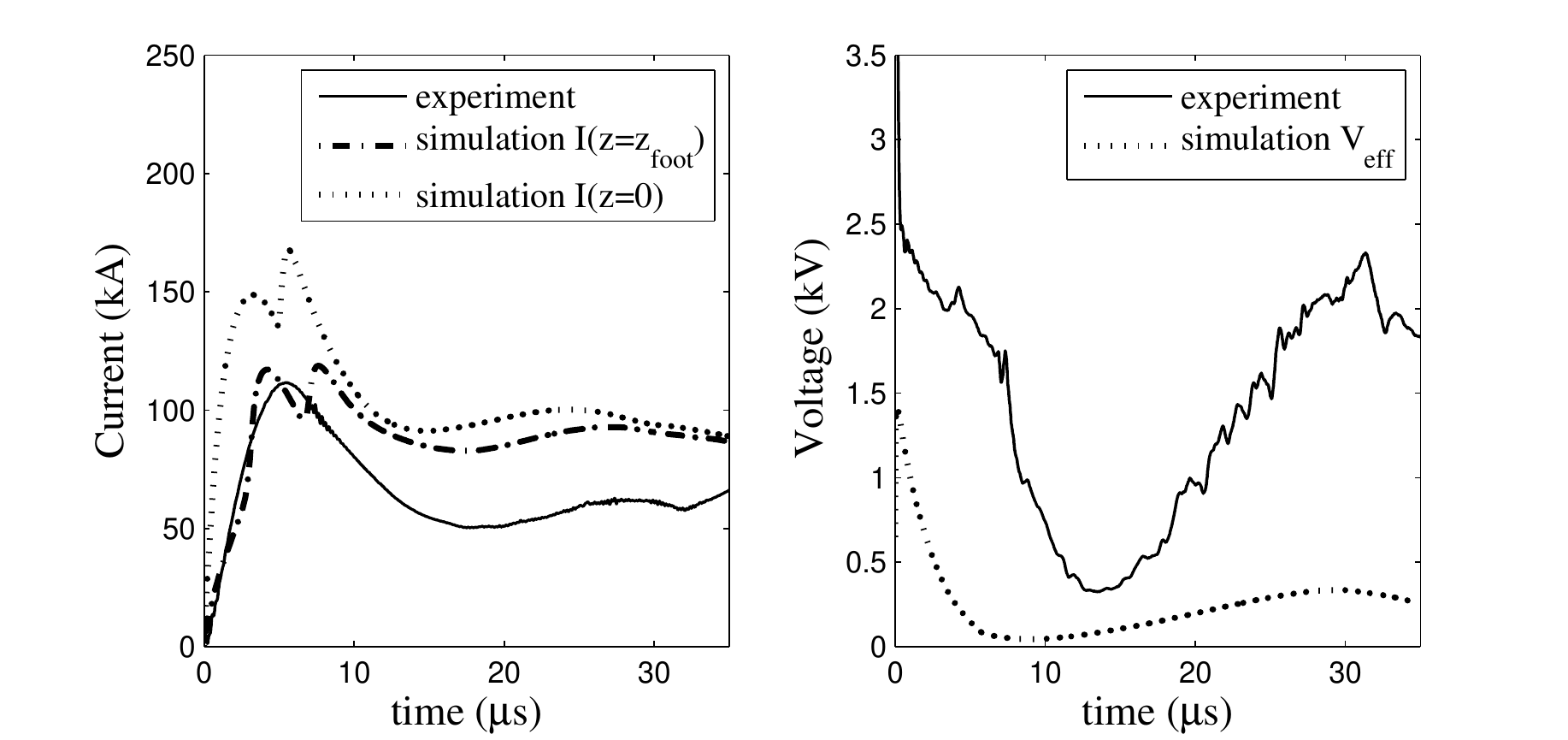} 
\caption{Left: evolution of the total positive poloidal current. 
Right: evolution of the (effective) voltage. The solid curves are measured 
in a typical argon plasma experiment (shot \# 12780, plasma discharged at $5$ kV). 
The dotted curves and dash-dot curves are simulation results calculated at $z=0$ 
and $z=z_{\rm foot}$ by Eq. \ref{eqn:eff_voltage}.}\label{fig:VI_compare}
\end{figure}

First, we examine the overall global energetics of the jet. 
The kinetic energy, magnetic energy and thermal energy in different regions are 
calculated by integrating dimensional quantities $\rho v^2/2$, $e_B=B^2/(2\mu_0)$ and $P_g/(\gamma-1)$ 
over the volume of interest for comparison with experiment. The evolution of these various types 
of energy are plotted in Fig. \ref{fig:energy_partition}.

The simulation starts with a finite thermal energy and a small 
magnetic energy from the initial poloidal magnetic field. During the first $5$ $\mu$s, 
the toroidal field is injected into the engine region at a very fast rate, leading to a 
rapid rise in total magnetic energy. Meanwhile, the injected toroidal field 
continuously applies a Lorentz force to the plasma, converting magnetic 
energy into kinetic energy. At $5$ $\mu$s, this energy conversion rate exceeds 
the declining toroidal field injection rate, and the magnetic energy of the entire 
simulation domain begins to drop. This dropping trend is terminated by the second 
fast injection occurring at later time. At $10$ $\mu$s, the relative amounts of 
magnetic and kinetic energy in the engine region reach a quasi-equilibrium 
state where magnetic energy dominates and remains roughly constant. 
However, the magnetic and kinetic energy in the jet region continue 
growing at constant rates. Therefore magnetic energy injected in the 
engine region is effectively transferred to the jet region because the 
energy in the engine region stays saturated. The  energy partition and 
evolution are consistent with estimation for the experiment\citep[see][Chapter 3]{Kumar_Thesis}.

The thermal energy is insignificant compared to the magnetic and 
kinetic energies. The thermal energy has a small rise in early time due 
to the adiabatic heating from the collimation, and then slowly decreases because of the 
mass loss at the domain boundaries. Heating during the jet evolution is in 
general also not important in the experiment.

In Section 3.3.2, we showed that the jet is accelerated by the plasma pressure gradient 
along the central axis. This pressure gradient is caused by the non-uniform toroidal 
field pinching. In the jet region, the rate of increase of kinetic energy greatly 
exceeds the decrease of the thermal energy. Therefore it is confirmed that the jet 
gains kinetic energy ultimately from magnetic energy, not from thermal energy, 
i.e., the jet is magnetically driven.

The total power input into the system is given by
\begin{equation}
P_{\rm tot}\equiv\iiint(\dot{e}_B+\dot{e}_K+\dot{e}_T)dV\label{eqn:power_input},
\end{equation}
where $e_B$, $e_K$ and $e_T$ are the magnetic, kinetic and thermal energy density.

If we ignore the energy loss due to the outflow mass at the solving domain boundaries, 
the energy conservation law states that the rate of change of total energy equals the 
energy injection rate associated with toroidal field injection, i.e.,
\begin{equation}
P_{\rm tot}=P_{\rm inj}\equiv \iiint \dot{e}_{\rm inj}dV\qquad \dot{e}_{\rm inj}\equiv \gamma_b(t)\mathbf{B}_{\rm tor}\cdot\mathbf{B}\label{eqn:power_injection}.
\end{equation}
According to the analysis in Section 3.3.1, the power injection mainly occurs in the engine region, i.e.,
\begin{equation}
P_{\rm inj}\approx \iiint_{|z|<z_{\rm foot}} \dot{e}_{\rm inj}dV\equiv P_{\rm inj,engine}.
\end{equation}

Due to energy saturation in the engine region, there is also
\begin{equation}
P_{\rm tot}\approx \iiint_{|z|\geq z_{\rm foot}} (\dot{e}_B+\dot{e}_K+\dot{e}_T)dV\equiv P_{\rm jet}\quad\text{at large }t.
\end{equation}

Therefore
\begin{equation}
P_{\rm inj,engine}\approx P_{\rm jet}\qquad\text{at large }t.
\end{equation}
This shows that the power input at the jet base is mainly used to accelerate the jet, and not for heating.

An effective voltage at the $z=0$ plane can be defined as
\begin{equation}
V_{\rm eff}\equiv \frac{P_{\rm tot}}{I(z=0)}\qquad I(z)=\iint_{J_z>0} J_zdxdy\label{eqn:eff_voltage}
\end{equation}
where $I(z)$ is the total positive poloidal current through the plane $z$.

Figure \ref{fig:VI_compare} shows that the poloidal current in the simulation is in good 
agreement with the experimental measurement. At $t\le3$ $\mu$s, the current at $z=z_{\rm foot}$ 
is less than $30\%$ of the current at the $z=0$ plane. This is because most of the toroidal injection 
occurs within the engine region. However, for $t>5$ $\mu$s, the current entering the jet region 
is comparable with the total current in the system, indicating that the engine region is 
injecting toroidal flux into the jet region.

It is difficult in the experiment to measure the voltage across the plasma precisely 
because the impedance of the plasma is very low. The voltage measurement 
given by the solid curve in Fig. \ref{fig:VI_compare} contains the plasma voltage 
drop as well as voltage drops on the cables and connectors. The effective voltage 
in the simulation is expected to be comparable to but lower than the experiment 
measurement, as is generally the case in Fig. \ref{fig:VI_compare}.

The global energy and electric characteristics comparison between the 
simulation and experiment confirm that the simulation captures the essential features. 
The jet is MHD-driven and gains kinetic energy from magnetic energy. In the 
following sections, we discuss the detailed process of jet collimation and 
propagation and various properties of the jet.

\subsubsection{Jet Collimation and Propagation}

\begin{figure}[t!]
\figurenum{4}
\epsscale{1}
\plotone{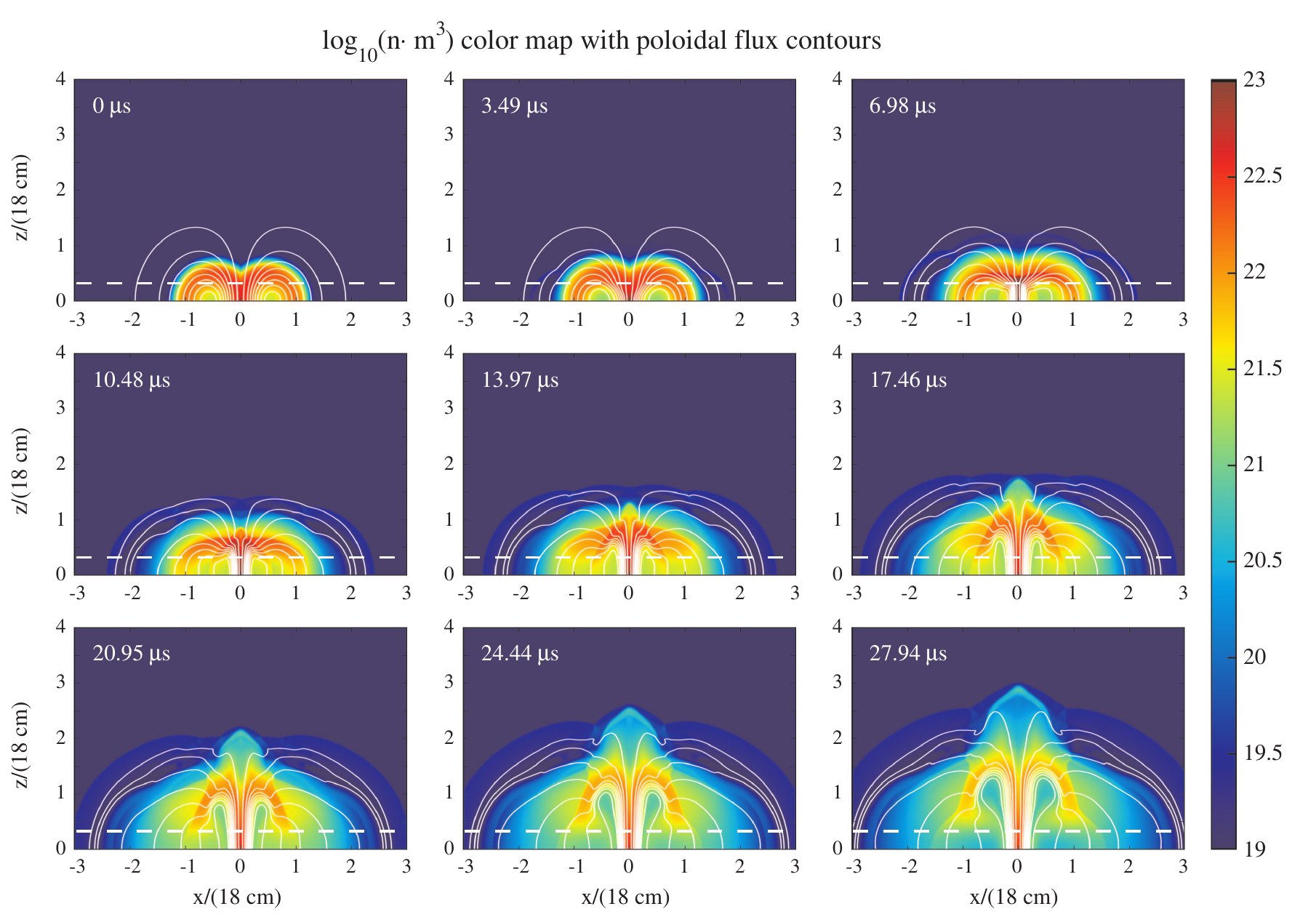} 
\caption{Evolution of the density distribution (color map) 
and azimuthally-averaged poloidal flux surfaces (white contours) in $xz$ plane ($z>0$) 
from $t=0$ to $0.48$ with $0.06$ interframe time, corresponding to dimensional time 
from $0$ to $27.94$ $\mu$s with interframe time $3.49$ $\mu$s. The color represents 
the common logarithm of the total particle number density $n=n_e+n_i$ in m$^{-3}$. 
Each frame contains 13 evenly spaced flux contours from $0.05$ mWb to $1.45$ mWb 
every $0.2$ mWb. The white horizontal dash lines in each frames mark the position of 
$z_{\rm foot}=0.307\Rightarrow5.5$ cm.
}\label{fig:den_evolution}
\end{figure}

\begin{figure}[h!]
\figurenum{5}
\epsscale{1}
\plotone{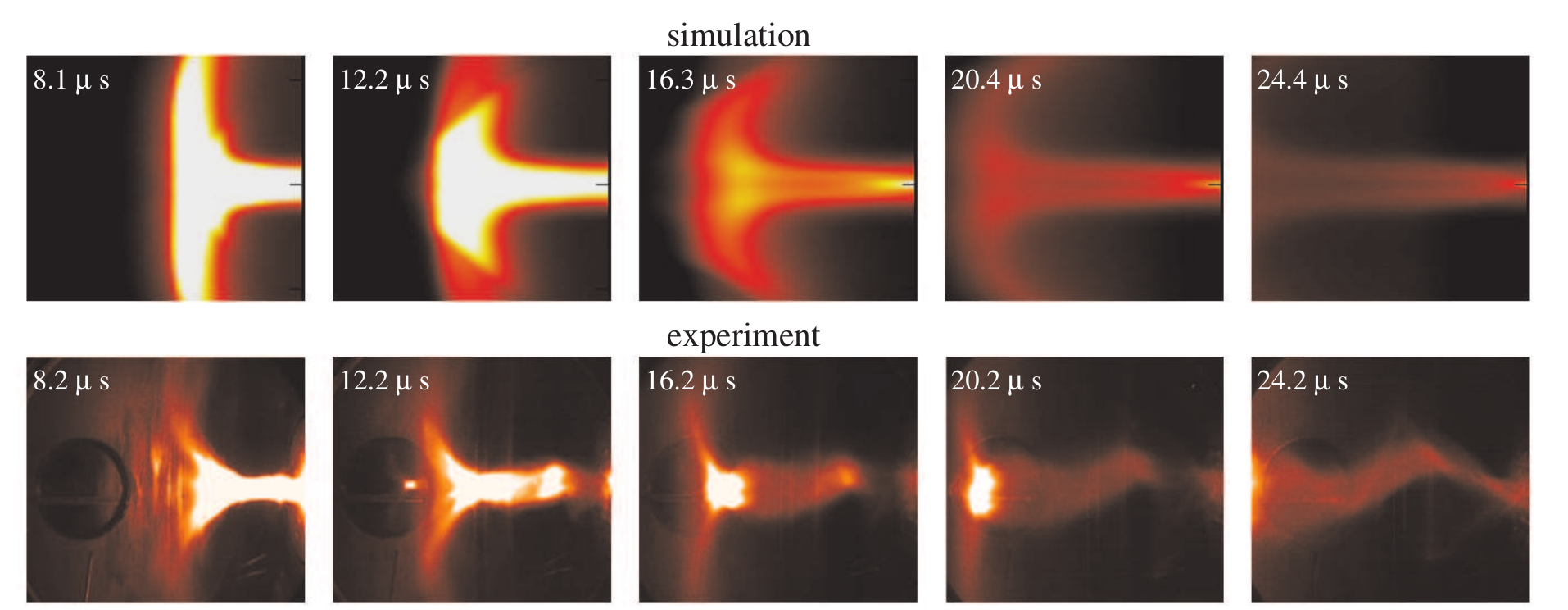} 
\caption{Top panels: distribution of line-of-sight integration of square of density in simulation, i.e., $\int n^2(x,y,z)dy$. Lower panels: false color images of a typical Argon jet experiment in visible band taken by IMACON 200 camera placed almost perpendicular to the jet axis (along $r$ direction). Shot \# 11082. The second frame also shows a reflected jet image on a glass window behind the jet. The plots are rotated $90^{\circ}$ about $(x=0,z=0)$, and are scaled to be $26$ cm in $z$ (horizontal) direction  by $22$ cm in $x$ (vertical) direction. The respective color tables for both the simulation and experimental images do not change with time.}\label{fig:emission}
\end{figure}

According to the analysis in Section 3 the $A=9$ localized toroidal field injection, 
quantified by Eq. (\ref{eqn:injection_coef}), will generate a pinch force that 
collimates the plasma near the $z=0$ plane. Meanwhile, the plasma pressure gradient 
along the axis, caused by the $z$ gradient of collimation force on the jet surfaces, 
will accelerate the plasma away from the $z=0$ plane.
The evolution of the 
plasma is given in Fig. \ref{fig:den_evolution} which presents the time sequence 
of plasma density in the $xz$ ($rz$) plane overlaid by azimuthally-averaged 
poloidal magnetic field contours. 
Figure \ref{fig:den_evolution} shows that plasma with frozen-in poloidal field is pinched 
radially and lengthened axially. Starting from a torus structure around the origin, 
the plasma eventually forms a dense collimated jet with a radius 
$r\simeq 0.2\Rightarrow3.6$ cm (at $z=0$) and height $h\simeq 2\Rightarrow36$ cm 
at $\sim 30$ $\mu$s. The radius-length ratio of the plasma decreases from 
$\approx 1:1$ to $\approx 1:10$. Consequentially, a more than five times 
amplification of density and poloidal field is observed to be associated with the 
collimation process in the simulation, consistent with the experimental 
measurement by \citet{Yun_You_Bellan_2007}. The jet radius
$r\simeq 0.2$ at $z=0$ in the simulation is found where plasma density drops below
$5\%$ of the central density $\rho(r=0,z)$. An unmagnetized hydrodynamic shock bounding the global structure forms in the numerical simulation and propagates outward, as a result of supersonic jet flow propagating into the finite pressure background; this shock is not observed in the experiment because of the lack of background plasma. Here we define the jet head as the leading edge of dense magnetized plasma along the central axis. This leading edge corresponds to the top of the $T$-shaped shell in Fig. \ref{fig:den_evolution} (from $z=0$ to $z\sim 1.8$ at $27.94$ $\mu$s, see also in Fig. \ref{fig:emission}). The jet head is the point where all the poloidal flux bends and returns back to the mid-plane. In front of the jet head, plasma is essentially unmagnetized and the density drops from $\sim 10^{22}$ m$^{-3}$ to $<10^{20}$ m$^{-3}$. Therefore the hydro shock and its downstream region from the $T$-shell to the shock front are not considered as part of the jet, but rather the termination of the entire global structure. Figure \ref{fig:den_evolution} also shows that the entire plasma structure remains axisymmetric in the simulation.

The high speed images of the experiment plasma jets shown in Fig. \ref{fig:expt} are integration of plasma atomic line emission along the line of sight. Generally atomic line emission is proportional to the square of density. Therefore we calculate the line-of-sight integration of density squared of the simulation jet and plot the equivalent ``emission'' images in Fig. \ref{fig:emission}, along with five experimental plasma images. The plasma is optically thin. Figure \ref{fig:emission} shows that simulation and experimental jets have similar radius, length/velocity, brightness variation and the relatively flat and bright plasma at jet head, a $T$-shaped structure. This $T$-shaped structure is a signature of return flux (also see the structure at the top of jet in Fig. \ref{fig:den_evolution}). Due to the lack of any background pressure, the experimental jet has a much flatter return flux structure, compared to the $T$-shaped structure shown in simulation images at later times. This structure is much dimmer in Fig. \ref{fig:expt} because for those figures the camera was not placed perpendicular to the jet so the line of sight does not lie entirely in the $T$-shell structure. Note that the experimental jet starts to kink at $20$ $\mu$s but the jet still propagates in a similar manner and remains attached to the center electrode. 

\begin{figure}[t!]
\figurenum{6}
\epsscale{1}
\plotone{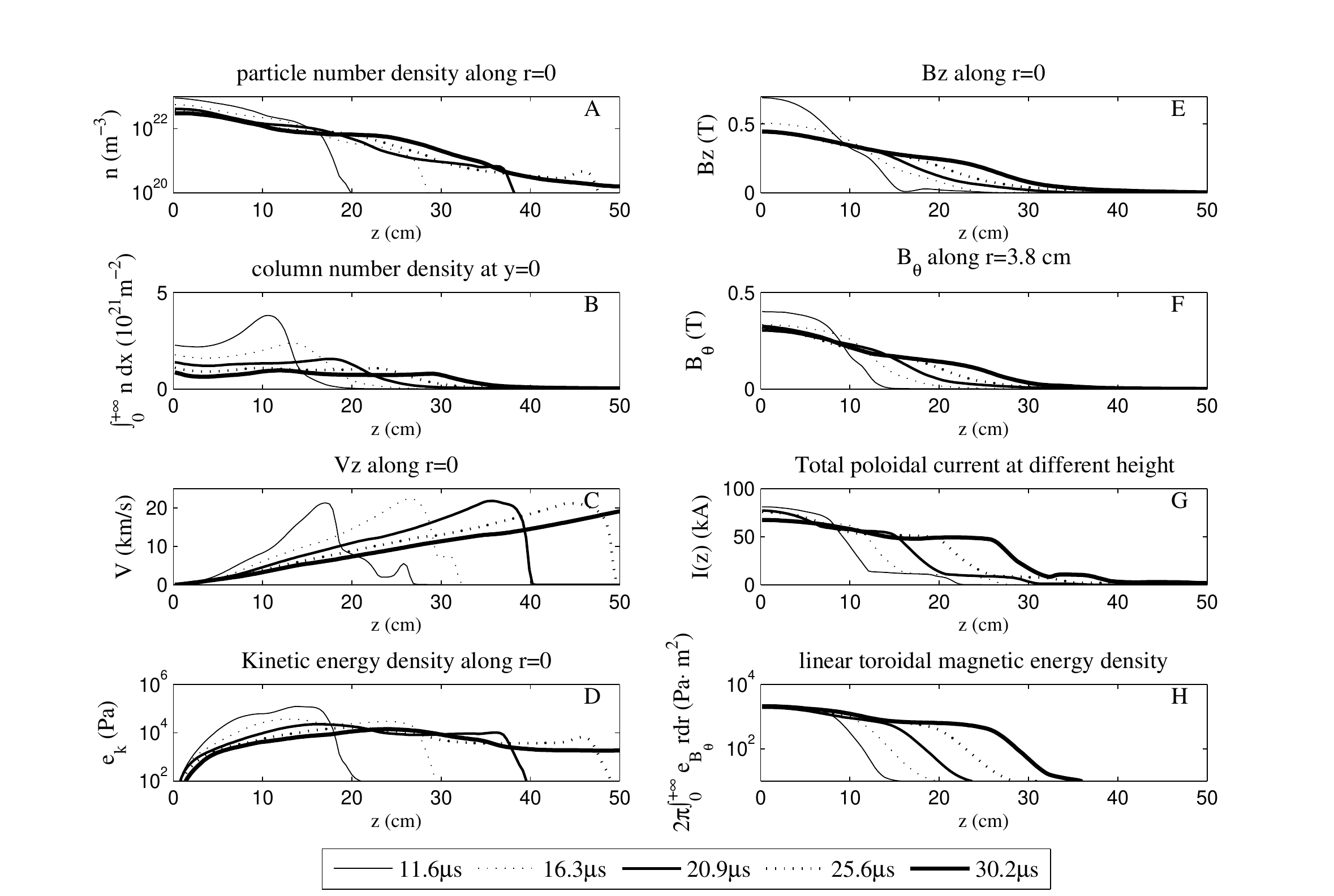} 
\caption{ Particle number density along central axis $n(r=0,z)$ (panel A), 
column particle number density $\int ndx$ along $y=0$ (panel B), axial velocity $v_z$ 
along $r=0$ (panel C), axial kinetic energy density $e_k=\rho v_z^2/2$ along $r=0$ (panel D), 
axial magnetic field $B_z$ along $r=0$ (panel E), toroidal magnetic field 
$B_{\theta}(r,z)$ at $r=0.21$ ($3.8$ cm) (panel F), total poloidal current $I(z)\equiv \max_r I(r,z)$, where $\mu_0(r,z)=B_{\theta}(r,z)/2\pi r$ (panel G), and total toroidal field energy at each height $\int_0^{\infty}e_{B_{\theta}} d\theta rdr$ (panel H) where $e_{B_{\theta}}=B_{\theta}^2/2\,u_0$.
}\label{fig:axial_profile}
\end{figure}

Although the localized toroidal field (poloidal current) injection is confined to the engine 
region ($|z|<z_{\rm foot}$, below the dashed lines in Fig. \ref{fig:den_evolution}), the plasma 
nevertheless collimates in the jet region. This is because the poloidal current, pre-injected 
in the engine region, propagates into the jet region along with the plasma motion and so provides a pinch force to collimate 
the plasma there (Fig. \ref{fig:VI_compare}, also see Fig. \ref{fig:jet_structure} in the next sub-section). 
Hence the toroidal field injection actually occurs in both the engine region and jet region.
The injection in the engine region is realized artificially by 
Eq. (\ref{eqn:induction}), a non-ideal process; 
the injection in the jet region is achieved through the $z=z_{\rm foot}$ 
plane associated with the plasma dynamics. 

The detailed axial profile of the collimated jet is given in Fig. \ref{fig:axial_profile}, which 
plots density, kinetic and magnetic profiles along the central $z$ axis spanning from 
$11.6$ $\mu$s to $30.2$ $\mu$s. Although the experimental jet already undergoes a kink instability as early as $\sim 20$ $\mu$s, the simulation results at late times can still be used to study the expansion of the length of the axis of the kinked experimental jet according to Fig. \ref{fig:emission}.

The left four panels A-D in Fig. \ref{fig:axial_profile} show the evolution 
of jet's kinetic properties. 
The number density plots (panels A and B) show that mass is rearranged to become 
more elongated and more evenly distributed along the jet. Since the total 
mass is conserved in the solving domain, consequentially, the density or column 
density decreases along the jet body when the jet gets longer. Panels C and D show the axial velocity and kinetic energy are gradually 
developed along the jet. The plasma axial velocity decreases in the lab frame because of the jet elongation. 
In fact, panel C indicates that the axial velocity approximately follows a self-similar 
profile $v_z(t,z)\propto z/t$. Detailed calculation finds that $tv_z/z$ approaches 
$1$ for $z>z_{\rm foot}$ at later time, i.e., $v_z\rightarrow z/t$. Therefore the acceleration 
in the frame of jet is  $dv_z/dt=\partial_t v_z+v_z\partial_z v_z=0$. This means 
that the jet has reached a dynamic steady state and the entire jet is elongating 
as a whole. However, it is crucial to point out that the $v_z\propto z/t$ behavior is 
only true at later times, when the injection rate varies very slowly. At early times 
when injection rate has a large variation, the jet velocity profile is expected to 
be very different from self-similar behavior, with density accumulation/attenuation 
in some parts of the jet and even internal shocks. At the jet head, plasma flow slows down in the moving frame of plasma, density accumulation always occurs (see panel B), which is also observed in experiments \citep{Yun_Bellan_2010}. This accumulation can be regarded as an indicator of jet head, e.g., 
$z\approx16$ cm at $t=16.3$ $\mu$s and $z\approx28$ cm at $t=25.6$ $\mu$s. This gives a jet speed of $v_z\approx 13$ km s$^{-1}$, consistent with the experiment (Fig. \ref{fig:emission}). 

The jet speed is faster than the background plasma sound speed $c_s=3.1$ km s$^{-1}$. The supersonic jet flow is expected to excite a hydro shock with speed $v_s=[(3\gamma-1)/(6\gamma-4)+\sqrt{(3\gamma-1)^2/(6\gamma-4)^2+c_s^2/v_z^2}]\cdot v_z\approx 18$ km s$^{-1}$ where the adiabatic constant is $\gamma=5/3$ \citep{Kulsrud_2005}. This is consistent with the simulation results in panel C. Under the strong shock approximation $v_z\gg c_s$, the shock speed is $v_s\approx [(3\gamma-1)/(3\gamma-2)]\cdot v_z$. In the experiment, although a hydro shock is also expected, it is not feasible to measure it because the background density is too low. \citet{AMoser_Thesis} and \citet{Moser_Bellan_2012} had a $v_z\approx 16$ km s$^{-1}$ argon experiment jet collide with a pre-injected hydrogen neutral cloud with density $n\sim 10^{19}-10^{20}$ m$^{-3}$, and observed a hydro shock in the cloud with a speed of $v_s\sim 25$ km$/$s. This satisfied the strong shock solution with $\gamma=7/5$ for neutral diatomic gas.

\citet[Fig. 15, 17]{Yun_Bellan_2010} measure the density and velocity profiles of a typical nitrogen jet using Stark broadening and Doppler effect. It is found that the experimental jet has a typical density $(0.5-1.0)\times10^{23}$ m$^{-3}$, and the density profiles behave very similarly to the argon simulation jet in aspects like mass distribution, time-dependent profile evolution, and density accumulation at the jet head, especially for the column number density (Fig. \ref{fig:axial_profile} panel B). The velocity profiles of the experiment nitrogen jet also show similar trends as Fig. \ref{fig:axial_profile} panel C, e.g., velocity behind the jet head slows down in lab frame and the jet head travels at a roughly constant speed. In the experiment, because there is negligible background density, the measurable plasma velocity reaches zero at the jet head. In the simulation, however, the axial velocity profiles are terminated by the hydro shock in front of the jet head. \citet{Yun_Bellan_2010} show a smaller density decrease of the jet in the experiment than in the simulation, due to the continuous mass injection into the plasma through the gas feeding holes on the electrodes \citep{Stenson_Bellan_2012}. Continuous mass injection is not included in the simulation in order to reduce complexity. This results in a larger density attenuation in the simulation as the jet propagates (panel A and B). It is important to point out that the experimental nitrogen jets and argon jets do not have exactly the same conditions, so the discussion here on nitrogen jet, while identifying similar trends, is not quantitative.

As the jet lengthens, axial magnetic field embedded in the plasma is also stretched out, 
resulting in a quasi-uniform magnetic density along the jet axis. This is clearly evident 
by noticing the $B_z$ evolution in Fig. \ref{fig:axial_profile} panel E. At $11.6$ $\mu$s, 
$B_z$ attenuates from $0.7$ T to $0.35$ T in $9.5$ cm, while at $30.2$ $\mu$s this 
$2$-fold decay occurs in a distance of $25$ cm $\approx 6$ jet radius. 
Hence the axial magnetic field is becoming more uniform. Panel F, G and H demonstrate
that toroidal magnetic field and poloidal current propagate
along the jet body and reach the same distance as does the plasma density, 
despite the fact that toroidal field/poloidal current is injected in the engine region at small $z$.
The jet is thus still being collimated by the toroidal field/poloidal current even 
though the jet is already far from the engine region. The total positive poloidal current (panel G) and total toroidal magnetic energy density (panel H) become quite uniform along the jet in later time. Panel G also clearly indicates the jet head location, where all poloidal current turn back and results in a sharp decrease in total positive poloidal current at the jet head. The locations of this sharp decrease is consistent with the location of density accumulation shown in panel B.

According to Fig. \ref{fig:axial_profile} here and Fig. 17 in \citet{Yun_Bellan_2010}, there is no distinct jet head in either simulation or experiment. After the main jet body, plasma density and other characteristics, such as poloidal flux and current, take significant distance to reach zero. The reason is again the lack of background pressure. In the jet-neutral cloud collision experiment \citep{AMoser_Thesis,Moser_Bellan_2012}, a sharper jet head with significant amplified density and magnetic field is observed.

Although panels E \& F show $B_z$ along the axis remains comparable with $B_{\theta}$ at the jet boundary, we will show in Section 4.1.3 that this result does not conflict with \citet{Lynden-Bell_1996, Lynden-Bell_2003, SherwinLyndenBell_2007} or \citet{Zavala_Taylor_2005}, in which an increasing pitch angle $B_{\theta}/B_z$ is expected tracing magnetic field lines along the jet.

\begin{figure}[h]
\figurenum{7}
\epsscale{0.8}
\plotone{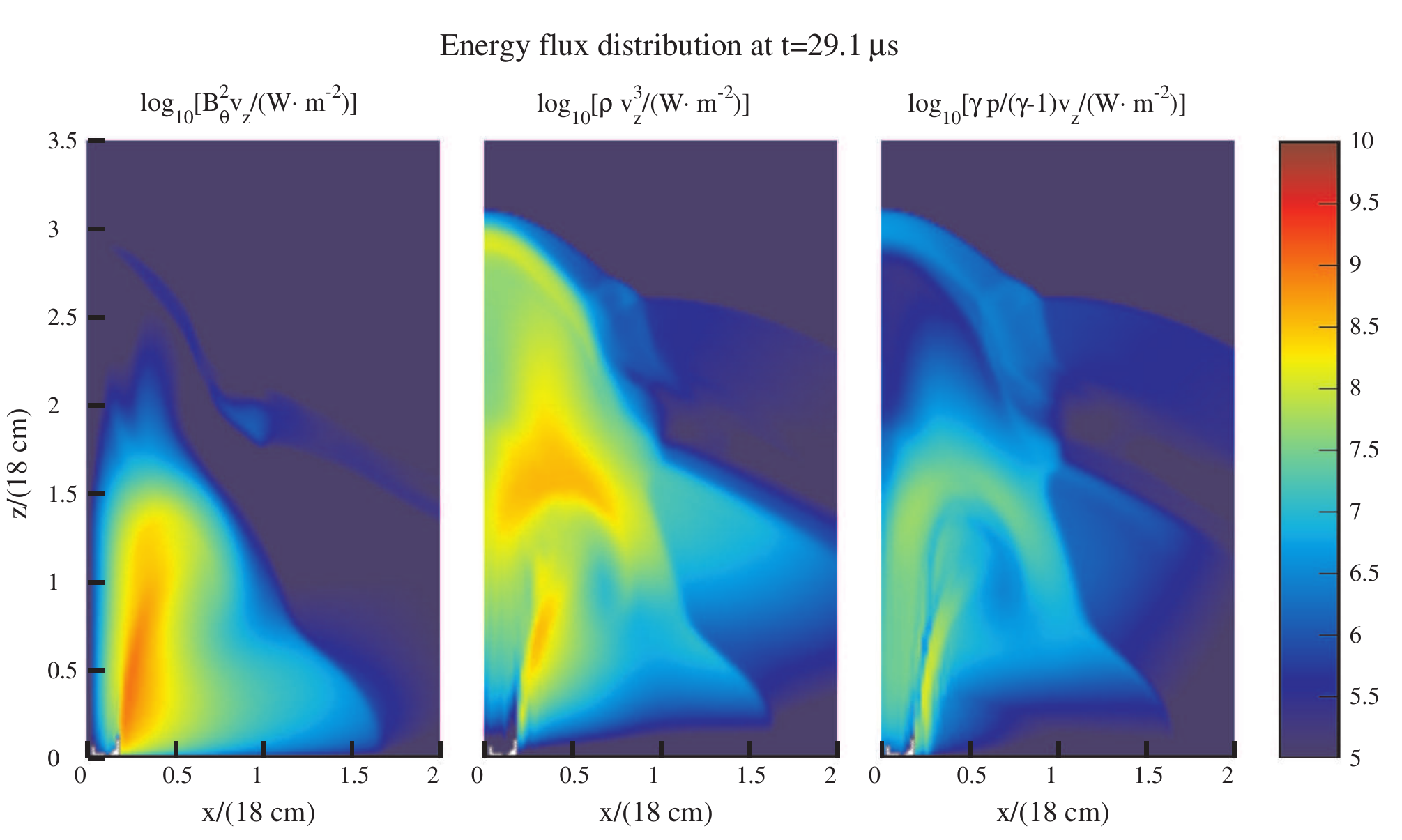} 
\caption{From left to right: distribution of logarithm of Poynting flux $\log_{10}(v_zB_{\theta}^2)$, kinetic flux $\log_{10}(\rho v_z^3)$ and enthalpy flux $\log_{10}(\gamma p/(\gamma-1)v_z)$ at $t=29.1$ $\mu$s. At this time, the jet head is at $z\approx 1.8$ or $32$ m and jet radius is about $r=0.2$ or $3.6$ cm. The SI unit for energy flux is W m$^{-2}$.} \label{fig:energy_flux}
\end{figure}

Figure \ref{fig:energy_flux} shows the distribution of  Poynting flux $B_{\theta}^2v_z$, kinetic flux $\rho v_z^3$ and enthalpy flux $\gamma p/(\gamma-1)v_z$ at $t=29.1$ $\mu$s. The figure shows that Poynting flux has successfully reached the height of jet head $z\approx1.8$, even though the toroidal field is injected at $z<0.307$. Poynting flux is generally $2-10$ times larger than kinetic flux, and two to three orders of magnitude larger than thermal flux, showing that the jet is MHD driven and magnetically dominated. However, at small radius where $B_{\theta}$ is small, kinetic and thermal flux are larger than Poynting flux. The hydro shock in front of the jet carries a notable amount of kinetic energy due to the fast expansion velocity.

\subsubsection{Jet Structure and the Global Magnetic Field Configuration}

We have shown that a collimated jet automatically forms in the jet region when 
toroidal field is injected into the engine region. We now examine the jet structure in the jet region.

\begin{figure}[t!]
\figurenum{8}
\epsscale{1}
\plotone{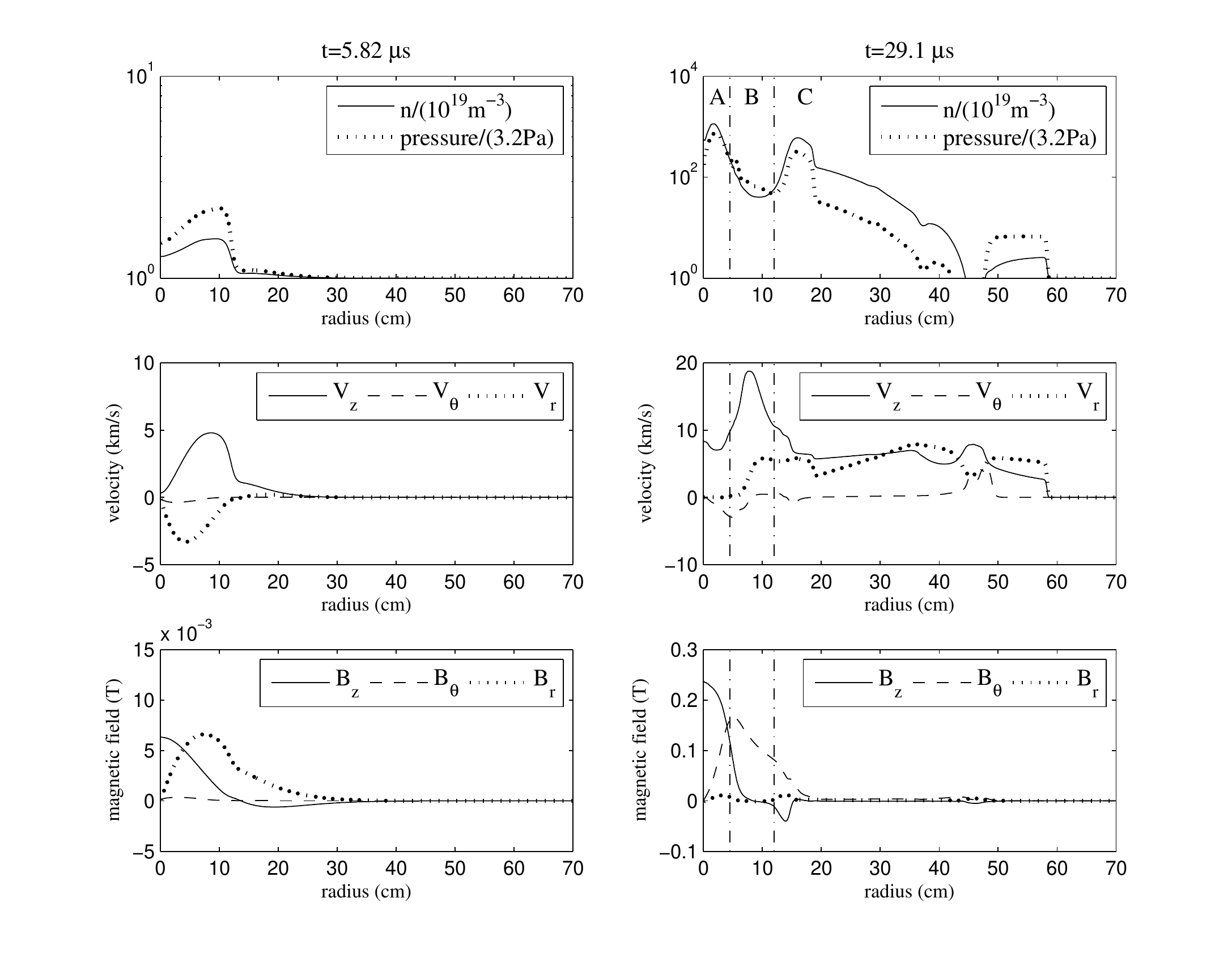} 
\caption{Radial profiles of the jet at $z=20.5$ cm ($15$ cm above $z_{foot}$ at 
$t=0.1$ or $5.82$ $\mu$s (left three panels) and $t=0.5$ or $29.1$ $\mu$s (right three panels). 
Top two panels: particle number density $n(r)$ (in $10^{19}$ m$^{-3}$, solid curves) 
and thermal pressure $p(r)$ (in $3.2$ Pa, dotted curves). Middle two panels: 
velocity profiles in km s$^{-1}$. $v_z(r)$ (solid curves), $v_{\theta}(r)$ (dashed curves) 
and $v_r(r)$ (dotted curves). Bottom two panels: magnetic field profiles in Tesla. 
$B_z(r)$ (solid curves), $B_{\theta}(r)$ (dashed curves) and $B_r(r)$ (dotted curves). Each of the right three panels ($29.1$ $\mu$s) is 
divided into three regions $A$, $B$ and $C$ separated by two vertical dot-dash lines at $r=4.5$ cm and 
$r=12$ cm. See Section 4.1.3 for details.}\label{fig:r_profiles}
\end{figure}

Figure \ref{fig:r_profiles} plots the radial profiles of the plasma density, pressure, 
velocity and magnetic field at $z=1.14\Rightarrow 20.5 $ cm ($15$ cm above the $z_{\rm foot}$ plane)
at different times. At $5.8$ $\mu$s, according to Fig. \ref{fig:den_evolution}, a collimated 
jet structure has not yet formed, and the injection in the engine region has caused little 
impact at $z=20.5$ cm. As expected, the left three panels of Fig. \ref{fig:r_profiles} reveal a 
low density ($\sim 10^{19}$ m$^{-3}$), low velocity and very weakly magnetized plasma structure. 
(Note that vertical scales for $5.82$ $\mu$s and $29.1$ $\mu$s are different in 
Fig. \ref{fig:r_profiles}). However, the negative radial velocity between $1$ and $10$ cm 
shows that the collimation has already started at this time. At $29.1$ $\mu$s, a collimated
jet in steady-state is expected at $z=20.5$ cm because the jet head has travelled beyond $20.5$ cm
according to Fig. \ref{fig:den_evolution}. The right three panels of Fig. \ref{fig:r_profiles}
show that the entire radial profile can be divided into three regions from small to large
radii, namely the central column (jet, region $A$), the diffuse pinch region (region $B$) and
the return flux region (region $C$) (see also discussions of these structures in 
\citet{Nakamura_2006} and \citet{Colgate_etal_2014}).

\paragraph{Central column}
For $r\lesssim 4-5$ cm, the central jet is characterized by a 
$\sim 10^{22}$ m$^{-3}$ high density, a $\sim 10$ km s$^{-1}$ quasi-uniform axial 
velocity and a $\sim 0.24$ T axial magnetic field. The radial velocity 
is zero, indicating that collimation is complete and a radially balanced $z$-pinch 
configuration is maintained. The toroidal magnetic field gradually increases 
from $r=0$ to $r\approx 5$ cm at a roughly constant slope, suggesting that the
central jet is filled by a roughly uniform current $J_z$. The zero $B_r$
additionally demonstrates that the magnetic field is well confined inside the jet.
At the jet boundary, density, pressure, axial magnetic field and current density
drop rapidly and connect to the diffuse pinch region. Specifically, at $r=5$ cm, the
plasma density is already less than $15\%$ of the maximal density $1.14\times10^{23}$
m$^{-3}$ at $r=1.7$ cm. The density dip at $r=0$ results from the initial torus-shaped
mass distribution.

\paragraph{Diffuse pinch region}
For $5$ cm$\lesssim r\lesssim 12$ cm, there is a relatively large region filled by 
low density plasma ($\sim 5\times 10^{20}$ m$^{-3}$) surrounding the central dense jet. 
The toroidal magnetic field $B_{\theta}$ scales as $r^{-0.96}\approx 1/r$ in this region, 
showing that the poloidal current is almost zero. Detailed calculation shows that $87\%$ of
total axial current $I_Z$ flows inside the central column $r<5$ cm, and another $13\%$ of
$I_Z$ exists in the $5$ cm$\lesssim r\lesssim 10$ cm region. The axial magnetic field $B_z$
drops to zero with a steep scaling $B_z\sim r^{-5.5}$ from $5$ cm to $8$ cm, and reverses 
polarity at $r=8.5$ cm. The radial magnetic field $B_r$ is $\lesssim 10^{-2}$ times 
weaker than $B_z$ and $B_{\theta}$. This region has a relatively fast axial velocity 
and finite radial velocity. However, because of the low density, the kinetic energy 
in this region is only $15\%$ of the toroidal magnetic energy in the same region, 
and is less than $10\%$ of the central column kinetic energy. Hence the diffuse 
pinch region is a toroidal magnetic field dominant region with low $J_z$. 

\paragraph{Return flux region}

Since the simulation starts with a complete global dipole magnetic field, 
the poloidal flux, carried by the central jet, must return to the central plane 
at some point. According to Fig. \ref{fig:den_evolution} and Fig. \ref{fig:axial_profile}, all the upward flux 
frozen into the dense plasma starts to return at the jet head. The return flux at 
$z=20.5$ cm is found in the narrow $12$ cm$\lesssim r \lesssim 15$ cm region 
and has a $\sim 0.04$ T negative strength. The toroidal field sharply decays 
to zero in this region as well, indicating the existence of a narrow return  
poloidal current sheet. The Lorentz force acting on this current sheet repels 
this region away from the central axis at a fast speed ($v_r\approx 6$ km s$^{-1}$), 
and piles up and compress plasma in $15$ cm$\lesssim r \lesssim 18$ cm and forms the $T$-shell shown in Fig. \ref{fig:den_evolution}.

The return flux region transitions to the background plasma configuration through a
hydrodynamic shock at $r\approx 50-60$ cm. At $t=29.2$ $\mu$s, since the return
flux region still has higher density and pressure compared to the background, the
unmagnetized shock expands radially at a supersonic velocity of $v_s\approx 6$ km s$^{-1}$
(sound speed $C_{s0}=3.1$ km s$^{-1}$, see Table \ref{tab:units}). At very late time, when
there is sufficient radial expansion, the density and pressure in the return flux
region are expected to be low enough so that the expansion will become sonic. The
entire jet structure is expected to transit to pressure confinement 
from inertial confinement \citep{Nakamura_2006}.

\begin{figure}[h]
\figurenum{9}
\epsscale{1}
\plotone{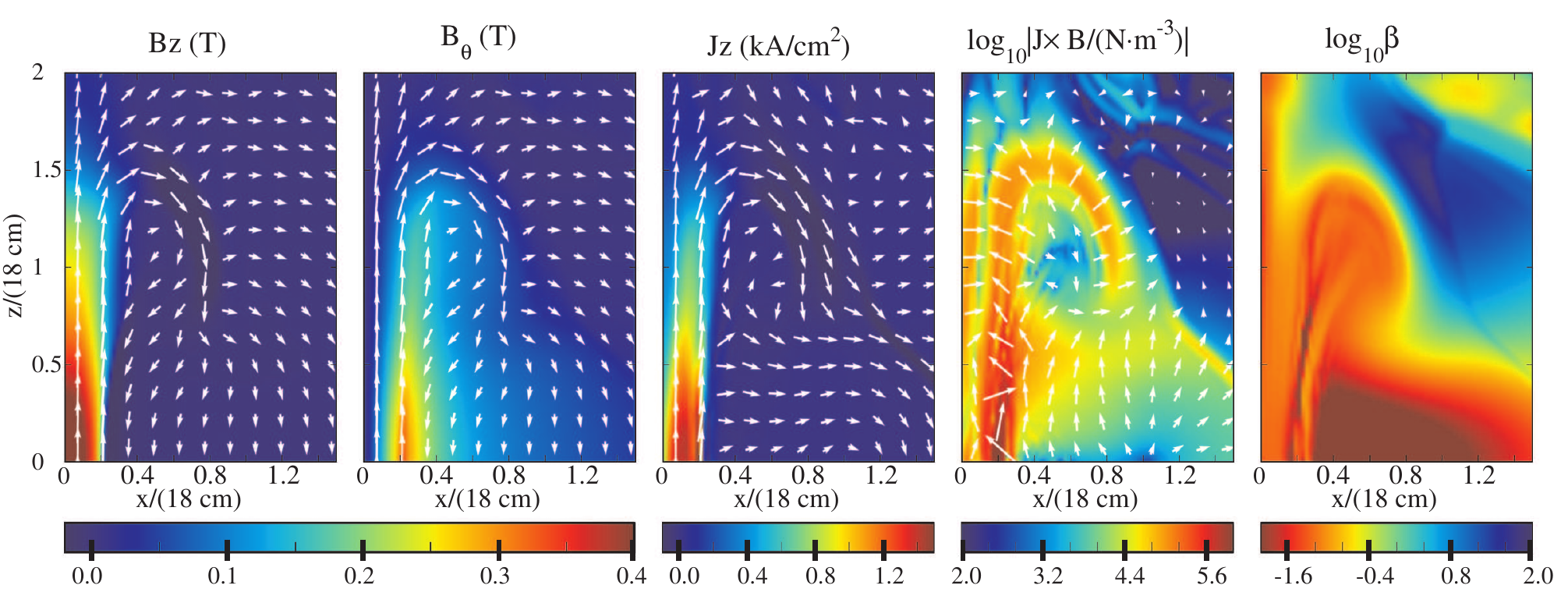} 
\caption{Cross-sectional view of plasma properties at $t=0.5$ or  $29.1$ $\mu$s. From left to right: axial magnetic field $B_z$ with poloidal field arrows, toroidal magnetic field $B_{\theta}$ with poloidal field arrows, axial current $J_z$ with poloidal current arrows, logarithm of Lorentz force density with poloidal $\mathbf{J}\times\mathbf{B}$ arrows, logarithm of plasma $\beta$ (ratio of thermal energy density to magnetic energy density) distribution. In all panels, the length of each arrow is proportional to the $1/5$ power of the corresponding quantity at the location of arrow center. For example, an arrow I with a half length of an arrow II means that the represented quantity at arrow I is only $1/2^5\sim 3\%$ of arrow II.}\label{fig:jet_structure}
\end{figure}

These radial profiles of the central jet confirm that the jet is highly magnetized and is 
MHD-collimated. The cross-sectional view of various plasma properties in Fig. \ref{fig:jet_structure} further validate this point. By comparing Fig. \ref{fig:jet_structure} with Fig. \ref{fig:den_evolution}, we find that the strong poloidal field and current 
are both confined in the dense plasma region (central jet region and the outer boundary 
of the return flux region). Poloidal field, current and toroidal field have been established from $z=0$ to $z=1.8$, same as the density and Poynting flux (Fig. \ref{fig:den_evolution} and \ref{fig:energy_flux}).

Figure \ref{fig:jet_structure} shows that the poloidal current is approximately parallel to the poloidal magnetic field in most of the region, especially in the central column, suggesting that the Lorentz force is dominantly poloidal, because the toroidal Lorentz force $\mathbf{F}_{\rm tor}=\mathbf{J}_{\rm pol}\times\mathbf{B}_{\rm pol}\approx 0$. This is consistent with the analysis given by Eq. (\ref{eqn:Ftor_Fr_ratio}). Detailed calculation finds that $F_{\rm tor}$ in the simulation is generally one to three orders of magnitude smaller than $F_{\rm pol}$. 
The Lorentz force distribution panel shows that $\mathbf{J}\times\mathbf{B}$ is extremely strong at the jet boundary especially at relatively low height. The Lorentz force at the jet boundary is radially 
inwards due to the self-pinch of the poloidal current, and is responsible for the collimation. The very large gradient of this pinching force along $z$ direction $\partial_z[(\mathbf{J}\times\mathbf{B})_r]$, equivalent to the gradient of toroidal magnetic energy $\partial_z(B_{\theta}^2)_r$, collimates the plasma gradually from lower $z$ to higher $z$, and ultimately accelerates the plasma. This demonstrates the MHD 
pumping mechanism in the current-driven plasma tube proposed by \citet{Bellan_2003} and verified in the Caltech plasma jet experiment \citep{Yun_You_Bellan_2007, Yun_Bellan_2010, Kumar_Bellan_2009}. Figure \ref{fig:jet_structure} also shows that the return flux/current are expanding outwards under a relatively strong Lorentz force. It is notable that at $z>0.7$ where the jet has not been fully collimated, the poloidal field is being compressed at very small radius, resulting in a radial outward Lorentz force.

The plasma $\beta$ panel shows that the central jet has a typical $\beta\approx 10^{-1.5}-10^{-1}$ ($0.03-0.1$), consistent with the experiment (Section 2). Hence the jet is magnetically dominated. 
The $\beta$ value is even smaller in the diffuse pinch region, due to the 
low plasma density and relatively strong toroidal magnetic field. The hydro shock 
has a very high $\beta$ value since it is essentially unmagnetized.


\begin{figure}[h]
\figurenum{10}
\epsscale{0.5}
\plotone{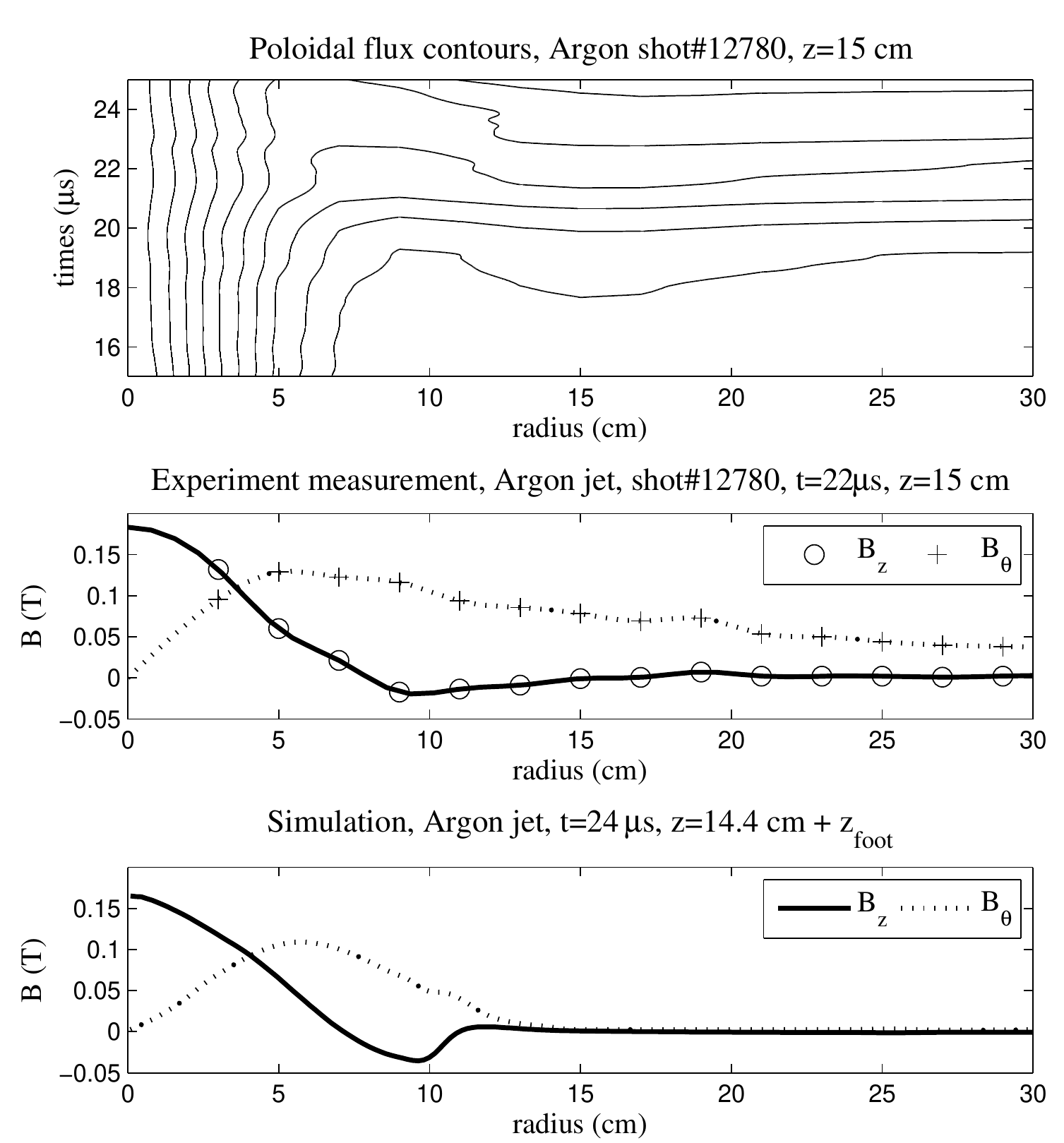} 
\caption{Top panel: poloidal magnetic field contours inferred from the MPA measurements from $t=15$ $\mu$s to $t=25$ $\mu$s. Mid and Bottom panels: magnetic field in axial (heavy solid curves) and azimuthal (dotted curves) direction measured in the experiment (mid panel) and in the simulation (bottom panel). The experimental measurements (top two panels) are obtained in argon jet experiment shot \# 12780. This experiment jet remains quasi-axisymmetric at $t=22$ $\mu$s. }\label{fig:magnetic_structure}
\end{figure}

Figure \ref{fig:magnetic_structure} compares the magnetic structure of the simulation jet 
with the experimental jet. The experimental measurements are obtained using the $1$ MHz 
20-channel MPA at $z=15$ cm from the electrode plane 
\citep{Romero_MPA} in a typical argon jet experiment. The top panel shows poloidal flux 
contours calculated from the MPA measurement from $t=15$ $\mu$s to $t=25$ $\mu$s, 
during which times the MPA has effectively ``scanned'' approximately $15$ cm distance along 
the $z$ direction in the moving frame of jet, although the MPA is fixed in the lab frame. 
The contours show that the magnetic field lines inside the jet ($r\lesssim 5$ cm) 
are quite collimated. The middle panel plots the radial profiles of $B_z$ and $B_{\theta}$ 
at $t=22$ $\mu$s in the experiment. The bottom panel gives the magnetic profiles in the 
simulation at $z=14.4$ cm$+z_{\rm foot}$ at $t=24$ $\mu$s. In both simulation and experiment, 
$B_z$ is $\simeq 0.2$ T at the central axis and reverses direction at $r\approx 7$ cm; 
$B_{\theta}$ rises quasi-linearly for small $r$ and peaks at $r=5$ cm. Hence $J_z$ is 
approximately constant within the central jet. Despite the excellent agreement in the 
central column region, it should be noted that the return current in the experiment 
extends to a much larger radius, leaving the entire $5$ cm$<r<30$ cm region devoid 
of current ($B_{\theta}\propto 1/r$). The return current in the simulation is at $\approx 8-15$ cm, where $B_{\theta}$ deviates from the $1/r$ behavior and quickly becomes zero. The return magnetic flux in the experiment, on the other hand, is located at $\approx 9-10$ cm, very similar to the simulation.


The $B_{\theta}$ due to the axial current in the jet produces a radially outward Lorentz force at the location of the return current. The expansion speed of the return current is determined by the density of the return flux region ($T$-shell in Fig. \ref{fig:den_evolution}) and the background pressure. The density of the return flux region $n\sim10^{21}$ m$^{-3}$ (Fig. \ref{fig:den_evolution} and \ref{fig:r_profiles}) is possibly too high compared to the experiment, although the experiment does not have accurate measurements of the low density plasma in the return current region. Also, the background pressure in the experiment ($10^{-7}$ torr $\sim 10^{-5}$ Pa for $n\sim 10^{15}$ m$^{-3}$ and $T=300$ K) is also much lower than the simulation background pressure ($p_0=3.2$ Pa for $n=10^{19}$ m$^{-3}$ and $T=2$ eV). Numerical investigation has found that the return current extends to a larger radius for a less dense $T$-shell or background. More discussion is given in Section 5.3 and 6.

\begin{figure}[h]
\figurenum{11}
\epsscale{0.9}
\plotone{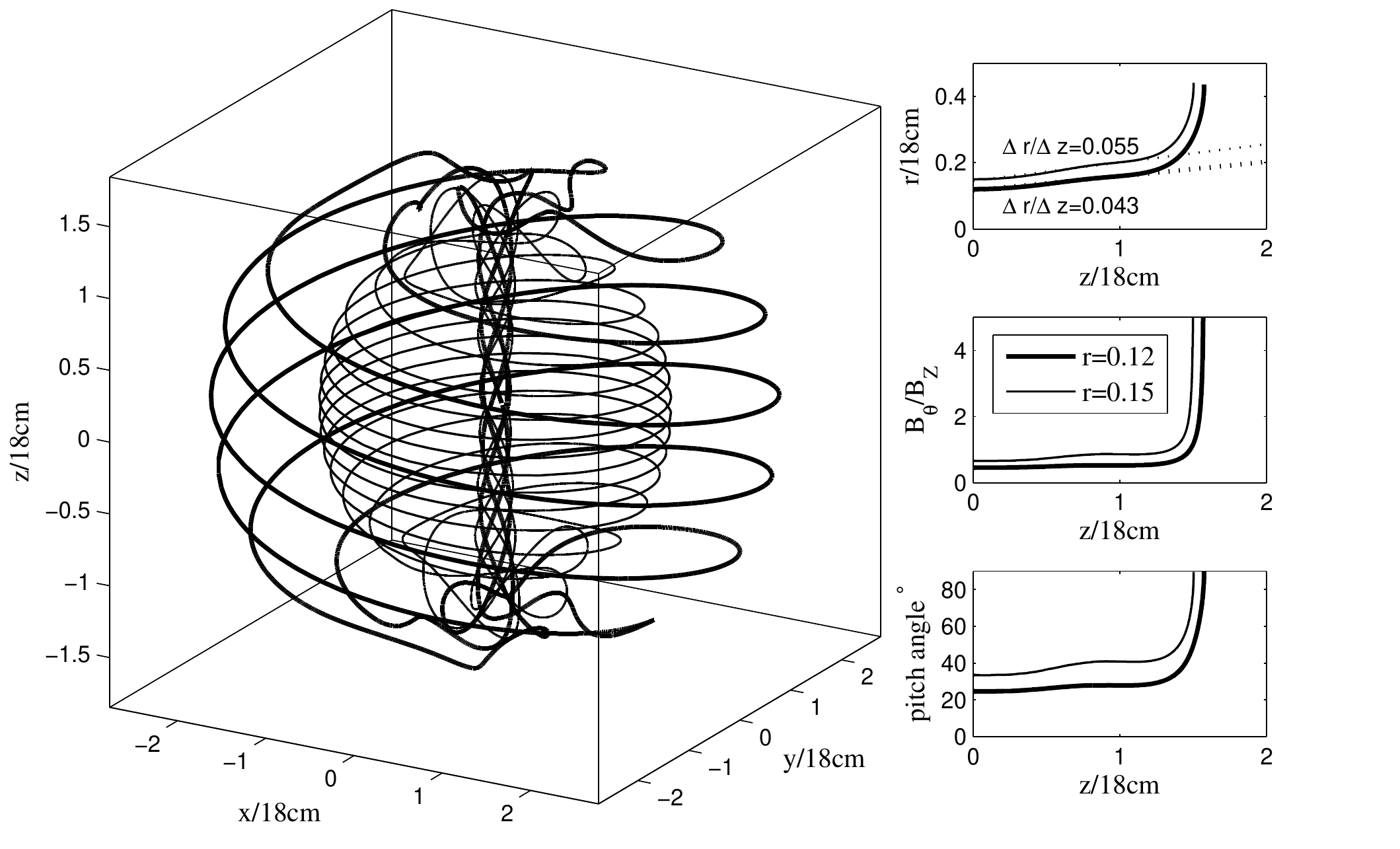} 
\caption{Left: 3D magnetic field structure of the simulation jet. The structure is composed by two groups of field lines starting from mid plane at $r=0.12$ ($2.16$ cm) and $r=0.15$ ($2.7$ cm). Each group contains four field lines azimuthally equally spaced starting at same radius. Upper right: radial location of each fieldline at different height. Linear regression within $0\le z\le 1.2$ gives $\Delta r/\Delta z=0.043$ with $R^2=0.981$ for field lines starting from $r=0.12$, and $\Delta r/\Delta z=0.055$ with $R^2=0.980$ for field lines starting from $r=0.15$. These correspond to opening angles $\theta=2\arctan(\Delta r/\Delta z)=4.9^{\circ}$ and $6.3^{\circ}$ for the two groups of field lines, respectively. Mid right: $B_{\theta}/B_z$ along the field line from mid plane to jet head. Lower right: the pitch of the magnetic field $\theta\equiv \arctan (B_{\theta}/B_z)$ in degree. In all three subplots, the thick curves represent fieldlines starting from $(r=0.12,z=0)$ and the thin curves represent fieldlines from $(r=0.18,z=0)$. The fieldlines are obtained at $t=0.5$ or $29.1\mu$s.}\label{fig:3d_magnetic_structure}
\end{figure}

Figure \ref{fig:3d_magnetic_structure} plots the 3D global magnetic field structure at $t=29.1$ $\mu$s, which shows a typical magnetic tower structure with upward flux along the jet and return flux surrounding the jet. The upward flux is twisted relative to the return flux. Tracing each field line from mid plane, the ratio $B_{\theta}/B_z$ is roughly constant along the central jet, and increases rapidly near the jet head because $B_z$ becomes zero at the turning point. Combining this figure with Fig. \ref{fig:axial_profile} panel E \& F, we find that at the jet head the poloidal field along the axis can remain comparable to the toroidal field at the jet boundary, although for each field line $B_{\theta}/B_z$ always increases. This is because the poloidal field and current do not bend over and return to mid plane at exactly the same height and same radius, i.e., there is no distinct jet head (also see Section 4.1.2). Both $B_z$ along the axis and $B_{\theta}$ at the jet boundary decrease gradually in the jet head region, giving a relatively constant ratio between them. The opening angles of the field lines shown in Fig. \ref{fig:3d_magnetic_structure} are $5-6^{\circ}$. Calculation shows that a field line starting from $r\sim 0.2$, essentially the boundary of the jet, has an opening angle of $11^{\circ}$; a field line from $r=0.1$ has an opening angle of $4^{\circ}$. It is found in the simulation that the opening angles become smaller as the toroidal field injection continuously accelerates and collimates the jet.

\subsubsection{Alfv\'en Velocity and Alfv\'en Surface}

\citet{Spruit_2010} categorizes the standard magnetocentrifugal 
acceleration model \citep[e.g.,][]{Blandford_1982} into three distinct 
regions: accretion disk, magnetic dominant region surrounding
the central objects and a distant kinetic dominant region. 
An Alfv\'en surface, on which the plasma velocity
equals the Alfv\'en velocity $v_A\equiv B/\sqrt{\mu_0\rho}$, 
separates the magnetic dominant region and
kinetic dominant region, since the ratio of plasma 
velocity to Alfv\'en velocity, $v/v_A=[(\rho v^2)/(B^2/\mu_0)]^{1/2}$, 
is the square root of the ratio of kinetic energy to magnetic energy. 

\begin{figure}[h]
\figurenum{12}
\epsscale{1}
\plotone{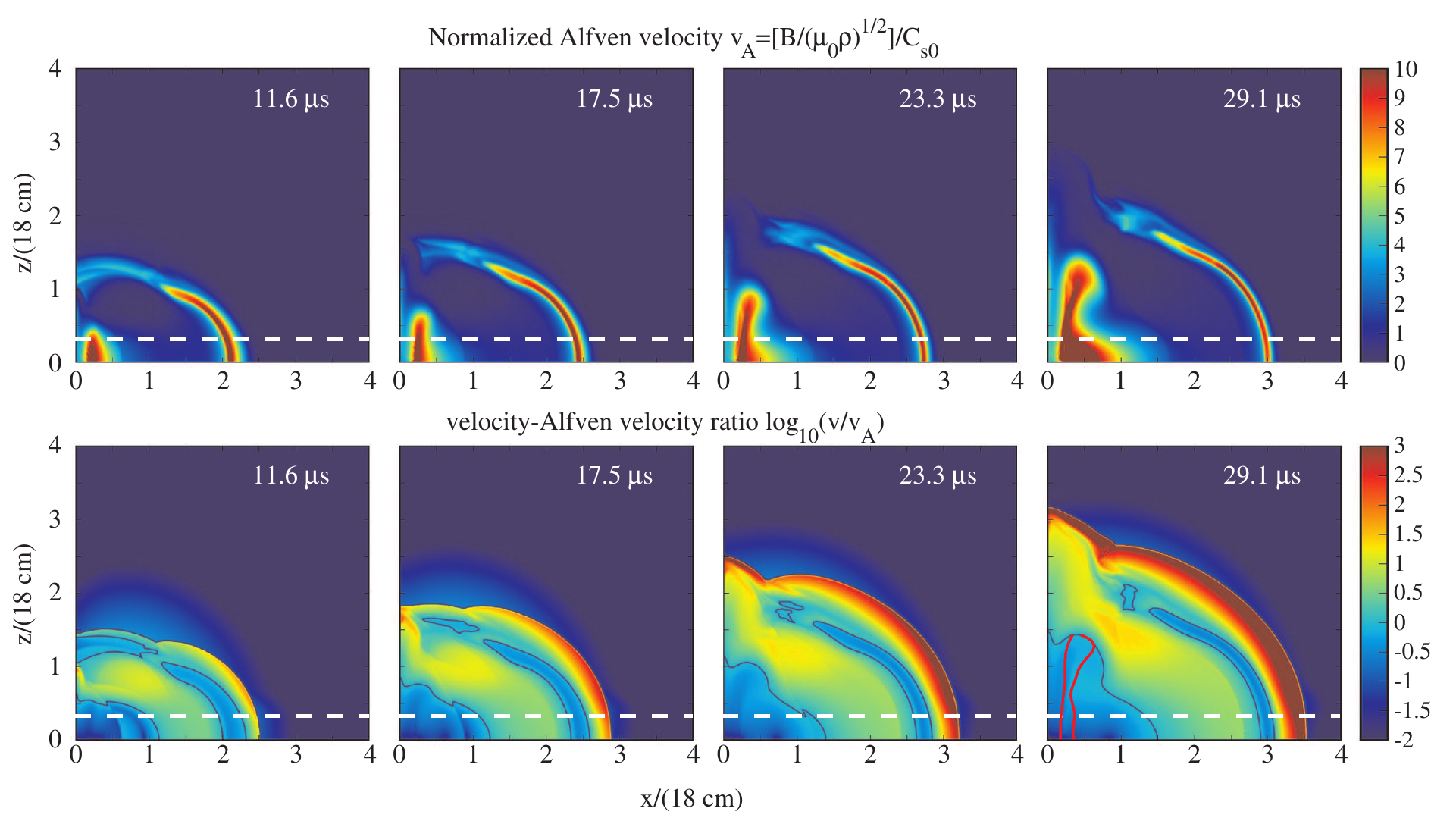} 
\caption{Alfv\'en velocity $v_A$ and velocity to Alfv\'en velocity ratio $v/v_A$ 
in $rz$ plane at different times. Top four panels: the color map of 
dimensionless Alfv\'en velocity $v_A/C_{s0}=[B/\sqrt{\mu_0\rho}]/C_{s0}$ 
($C_{s0}$ given in Table \ref{tab:units}). Bottom four panels: 
the color map of $\log_{10}(v/v_A)$ with $v=v_A$ contours 
(blue curves). The lower right panel is overlaid by two red curves. The one at smaller radius from $z=0$ to $z=1.6$ represents the contour of maximum
$B_{\theta}$ at each height, and is the boundary between the central jet region
and diffuse pinch region. The red curve at larger radius is the $J_z=0$ contour, that
separates the diffuse pinch region and the return flux region.}\label{fig:v_va}
\end{figure}

Figure \ref{fig:v_va} plots the distribution of dimensionless 
Alfv\'en velocity (top four panels) and
$v/v_A$ ratio (bottom four panels) in the $rz$ plane at different times. 
The boundaries of the central jet region and the diffuse
pinch region are overlaid on the lower right panel.
The figure shows that $v_A$ is
always high in the diffuse pinch region due to the low density 
and strong toroidal field.
In the central jet, $v_A$ remains roughly constant because of 
the quasi-constant density
and magnetic field configuration. The high Alfv\'en velocity 
region increases in volume together with the jet propagation.

The $v/v_A$ distribution plots show that the Alfv\'en surface, 
denoted by the innermost $v=v_A$ contour
curve, is also expanding. In the $+z$ direction, 
the Alfv\'en surface propagates from $0.5R_0=9$ cm at
$t=11.6$ $\mu$s to $1.5R_0=27$ cm at 
$t=29.1$ $\mu$s at a speed of $\approx 10$ km s$^{-1}$, similar
to the jet propagation speed. Along the central axis, the $v/v_A$ ratio 
gradually increases from $\ll 1$ at jet base
to $\sim 1$ at jet head, and becomes $\gg1$ at the hydro shock which has no magnetic field. According to Fig. \ref{fig:den_evolution}, 
\ref{fig:axial_profile} and \ref{fig:jet_structure}, the magnetic tower, wherein dense plasma encloses strong axial
magnetic field $B_z$ and axial current $J_z$, is inside the Alfv\'en surface. 
We point out here that the entire jet collimation and propagation dynamics 
is an integrated process. It is inappropriate to
characterize the jet as a hydrodynamic jet or magnetized jet 
simply based on the local $v/v_A$ ratio,
because the Alfv\'en surface is also expanding. Although the kinetic
energy of the global system extends beyond the Alfv\'en
surface in Fig. \ref{fig:v_va}, the magnetic tower is still an MHD driven jet. 
Outside the Alfv\'en surface,
according to Fig. \ref{fig:axial_profile}, both the poloidal and toroidal 
components of the magnetic field
decrease rapidly. 
The entire diffuse pinch region always has a relatively low $v/v_A$ ratio. 
Outside the Alfv\'en surface, there is another $v_A=v$ contour expanding outwards, which indicates the hydrodynamic shock. This is essentially the boundary of the entire large-scale jet structure. Outside this structure, both $v$ and $v_A$ are zero.

\subsection{Bernoulli Equation in MHD Driven Flow}
We have shown in detail the process of jet collimation and propagation 
resulting from the MHD mechanism. In Section 4.1.2, we have demonstrated that the 
jet gains its kinetic energy from magnetic energy; kinetic energy dominates near the 
jet head while magnetic energy dominates near the jet base. This has been 
quantitatively verified in the experiment.

Assuming that the Lorentz force balances the thermal pressure gradient in the 
radial direction, an axisymmetric model was proposed by \citet{Kumar_Bellan_2009} 
and \citet{Kumar_Thesis} to study the non-equilibrium steady-state flow along the 
axial direction. The model claims that a Bernoulli-like quantity involving the toroidal 
magnetic energy remains constant along the jet, i.e.,
\begin{equation}\label{eqn:MHDBernoulli}
\frac{\partial}{\partial z}\left[\rho v_z^2+\frac{B_{\theta,a}^2}{\mu_0}\left(1-\frac{r^2}{2a^2}\right)\right]=0
\end{equation}
where $a$ is the jet radius and $B_{\theta,a}=\mu_0I/(2\pi a)$ is the toroidal field 
strength at the jet boundary. Evaluating the expression at $r=0$ gives
\begin{equation}\label{eqn:MHDBernoulli_II}
\rho v_z^2+\frac{B_{\theta,a}^2}{\mu_0}=\rho v_z^2+\frac{\mu_0 I^2}{4\pi a^2}=\rm const,
\end{equation}
which is a Bernoulli-like equation. At $z\sim0$, the axial velocity $v_z\approx0$ 
so the magnetic energy dominates. At the jet head, $B_{\theta,a}\approx 0$ 
so the kinetic energy dominates. This is consistent with the analysis in Section 4.1.2. 
Evaluating Eq. (\ref{eqn:MHDBernoulli_II}) at $z=0$ and at the jet head yields
\begin{equation}\label{eqn:jet_velocity}
v_z|_{\text{jet head}}\simeq\left.\frac{I}{2\pi a}\sqrt{\frac{\mu_0}{\rho}}\right|_{z=0}\propto \frac{I}{\sqrt{\rho}}.
\end{equation}

\citet{Kumar_Bellan_2009} and \citet{Kumar_Thesis} report quantitative experimental 
measurements and show that the axial velocity of the MHD driven plasma jet is 
linearly proportional to the poloidal current, and inversely proportional to the 
square root of the jet density. Therefore Eq. \ref{eqn:jet_velocity}, a direct corollary 
of Eq. \ref{eqn:MHDBernoulli}, has been verified by the experiment.

Eqn. \ref{eqn:jet_velocity} can be understood from a semi-quantitative analysis. Since the injected Poynting flux or toroidal magnetic field energy will ultimately be used to accelerate the jet, an energy equal-partition gives $B_{\theta}^2\sim \rho v_z^2$. Hence $v_z\sim B_{\theta}/\sqrt{\rho}\sim I/\sqrt{\rho}$. Similar analysis and scaling can also be found in \citet{Lynden-Bell_1996, Lynden-Bell_2003, Uzdensky_MacFadyen_2006,Hennebelle_Fromang_2008}.

We now use the simulation to investigate this relation.

\subsubsection{Jet Velocity Dependence on the Poloidal Current}
\begin{figure}[h]
\figurenum{13}
\epsscale{0.5}
\plotone{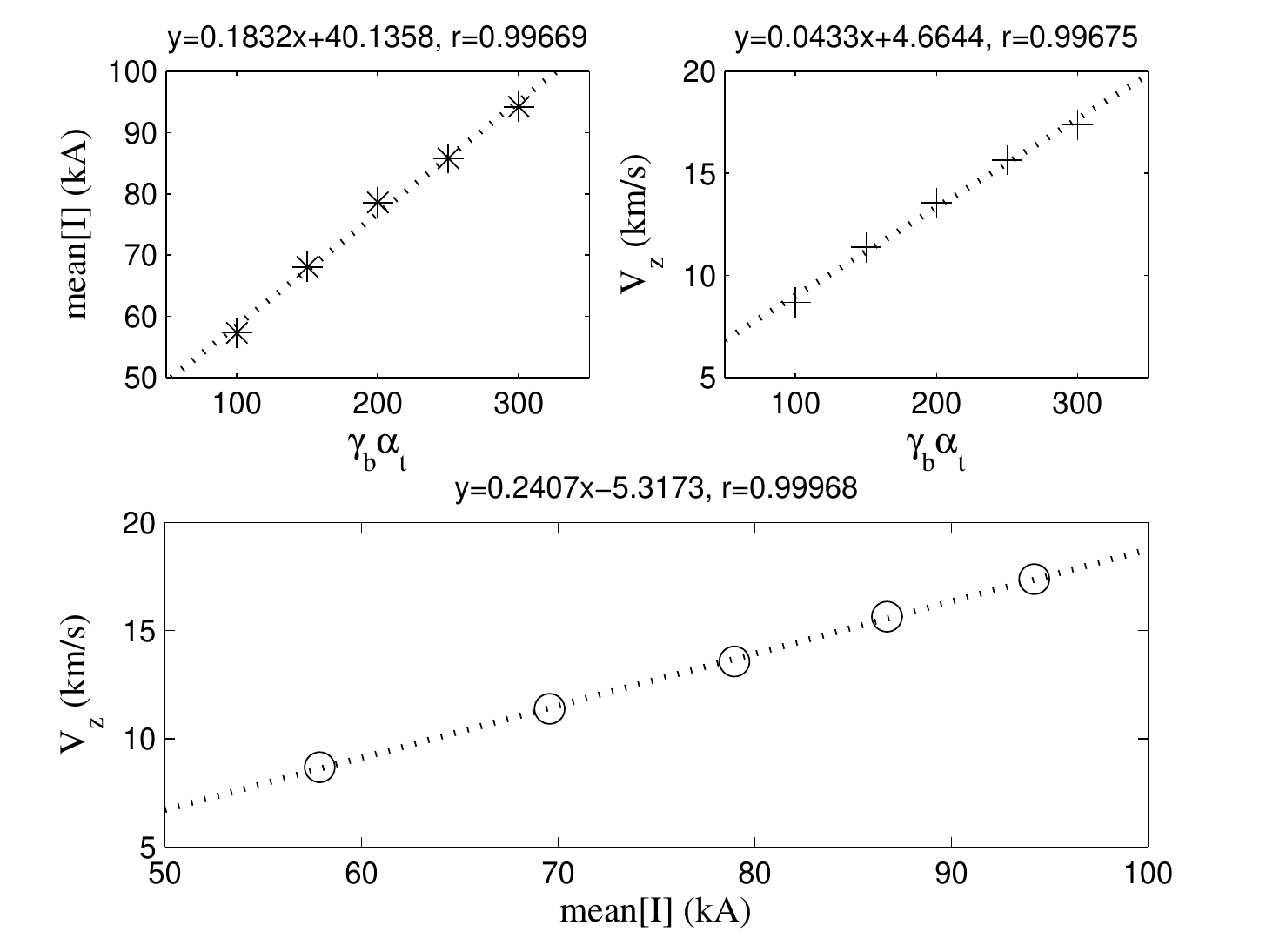} 
\caption{Jet velocity dependence in the simulation. Upper left: time-averaged 
total poloidal current for different injection rate $\gamma_b\alpha_t$ (asterisk symbols). 
Upper right: the averaged jet velocity at different injection rate 
$\gamma_b\alpha_t$ (plus symbols). Bottom: jet velocity vs. total poloidal current 
(open circle symbols). The averaging period is the time the jet head spends traveling 
from $z=30$ cm to $60$ cm. For each subplot, linear regression is performed 
(dotted lines) and the results are presented as the title.}\label{fig:jet_velocity}
\end{figure}

We use the same initial conditions as in Section 4.1, and the same localized toroidal field injection 
with the localization factor $A=9$. However, in order to control the total poloidal current, 
we use constant injection rates $\gamma_b\alpha_t$ throughout the simulation. Five simulations are performed with different time-independent injection rates 
over a wide range: $\gamma_b\alpha_t=100$, $150$, $200$, $250$ and $300$. The average jet velocity is computed using the time 
the jet head takes to travel from $z=30$ cm to $60$ cm 
($z=1.67$ to $3.33$ in reduced units). Here we define the location of the jet head as being where the plasma density drops below
$10^{21}$ m$^{-3}$ along the $z$ axis. According to Fig. \ref{fig:den_evolution} and Fig. \ref{fig:axial_profile}, this definition gives a sufficiently consistent estimation of the jet head location. The total poloidal current is also averaged 
over the same period. Figure \ref{fig:jet_velocity} shows that both the jet velocity and 
the time-averaged total poloidal current are proportional to the toroidal field injection rate 
$\gamma_b\alpha_t$. Thus the jet velocity is indeed proportional to the poloidal current.

\subsubsection{Jet Velocity Dependence on the Jet Density}
\citet{Kumar_Bellan_2009} and \citet{Kumar_Thesis} find that under the same 
experimental configuration, a deuterium plasma jet always propagates at a 
speed $=0.73\approx 1/\sqrt{2}$ times the speed of a hydrogen plasma jet. 
Hence $v_z\propto 1/\sqrt{\mu}\sim 1/\sqrt{\rho}$ is verified. In the simulation, 
this dependence is already incorporated by the normalization process in Section 3.1. 
Note that the simulation time unit is defined as
\begin{equation}
t_0\equiv\frac{R_0}{C_{s0}}\propto\frac{1}{C_{s0}}
\end{equation}
and
\begin{equation}
C_{s0}^2\propto\frac{1}{m_i}\propto\frac{1}{\mu},\qquad \mu\equiv \frac{m_i}{m_H},
\end{equation}
so the simulation time unit is proportional to $\sqrt{\mu}$. 
Therefore the simulation velocity unit is proportional to $1/\sqrt{\rho}$. 

Given that $n\approx 10^{22}$ m$^{-3}$ and $a\approx 4$ cm, 
Eq. (\ref{eqn:jet_velocity}) predicts 
$v_z/I\simeq\sqrt{\mu_0/\rho_0}/(2\pi a)=0.244$ m$/($s$\cdot$A$)
=0.244$ km$\cdot$s$^{-1}/$kA, which is consistent with the linear regression 
results given in the bottom panel of Fig. \ref{fig:jet_velocity}.

\subsubsection{A Direct Illustration of MHD Bernoulli Equation}
In fact, Eq. (\ref{eqn:MHDBernoulli}) can be easily verified directly by the simulation. 
Evaluating the equation at the jet radius $r=a$ gives
\begin{equation}
\frac{\partial}{\partial z}\left(\rho v_z^2+\frac{B_{\theta,a}^2}{2\mu_0}\right)=0
\Rightarrow\quad (e_k+e_{B_{\rm tor}}/2)|_{\text{jet radius}}=\rm const,\label{eqn:modified_MHD_Bernoulli}
\end{equation}
where the kinetic energy density is $e_k\equiv\rho v_z^2/2$ and 
the toroidal magnetic field energy density is $e_{B_{tor}}\equiv B_{\theta}^2/2\mu_0$.

\begin{figure}
\figurenum{14}
\epsscale{0.45}
\plotone{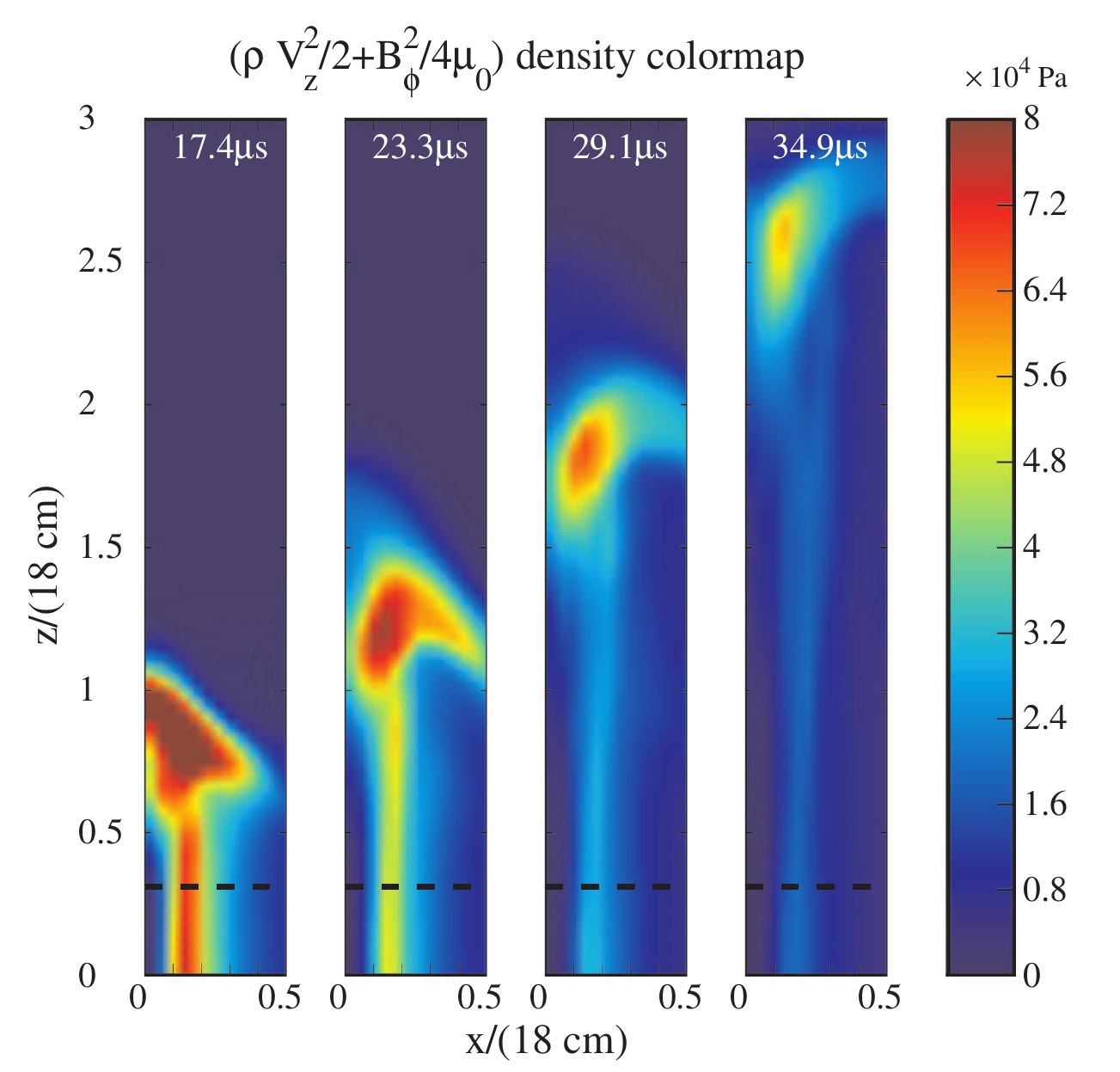} 
\plotone{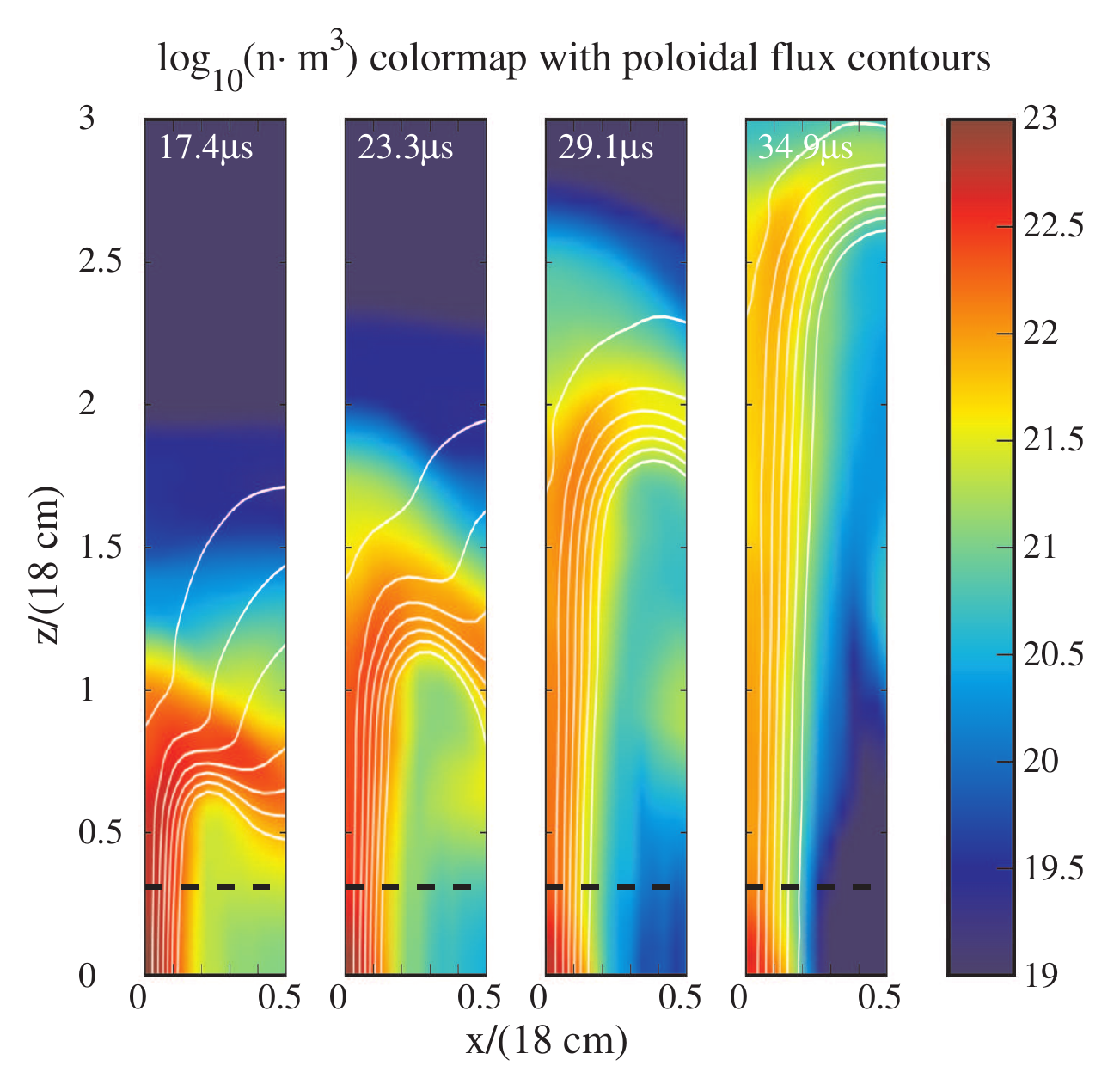} 
\plotone{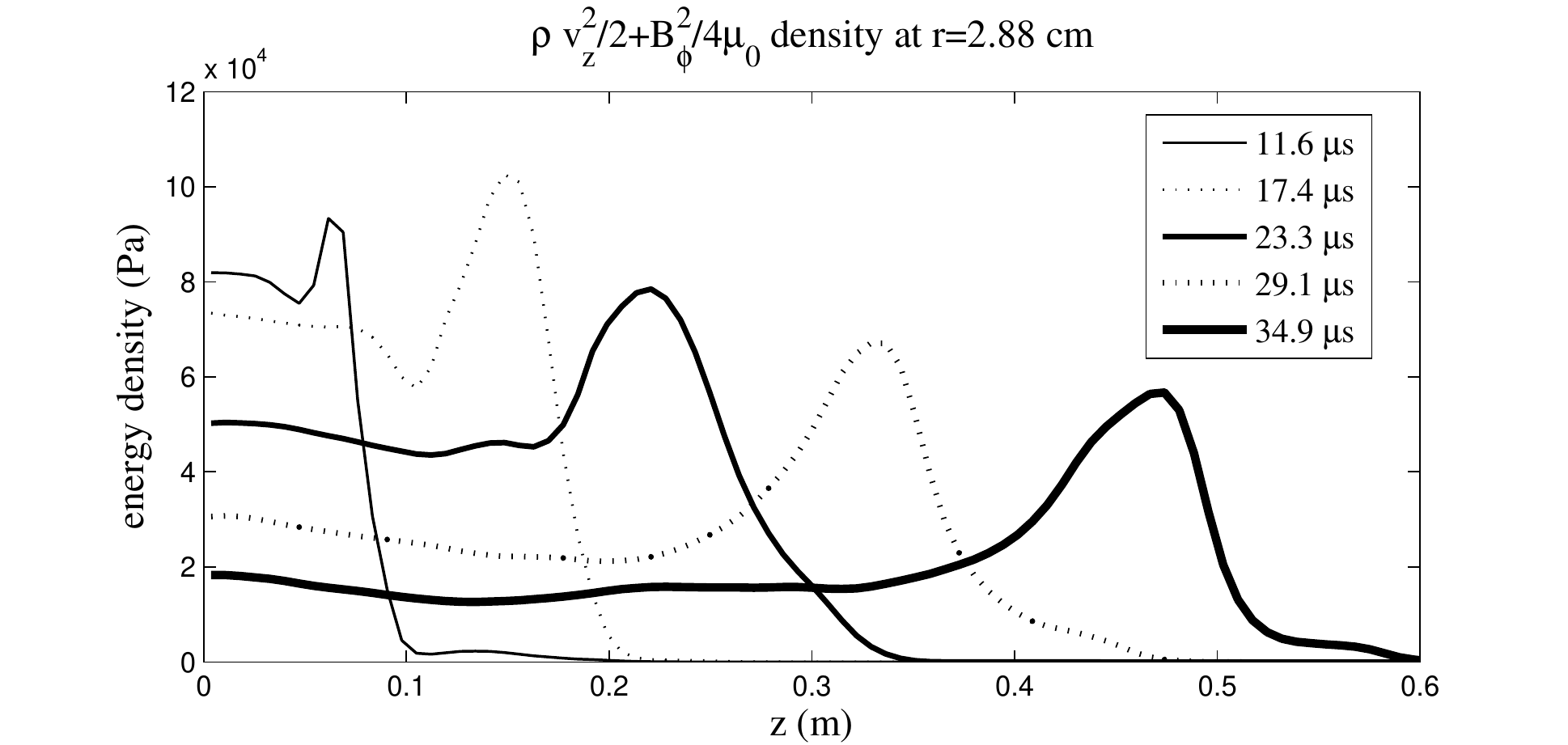} 
\caption{Top left panel: cross-sectional view of 
$(e_k+e_{B_{\rm tor}}/2)$ energy density on the $xz$ plane ($y=0$) from $17.4$ $\mu$s to $34.9$ $\mu$s. Top right panel: cross-sectional view of density distribution (log scale) at the same times as used for the plots in the top panel. Azimuthally averaged poloidal flux contours are overlaid. Note that the jet radius to length ratio has dropped to $\sim 1:20$ at 
$t=34.9$ $\mu$s. Bottom panel: $(e_k+e_{B_{\rm tor}}/2)$ at $r=2.88$ cm along the $z$ direction at different times. The plots are generated from the $\gamma_b\alpha_t=200$ simulation in Section 4.2.1.}\label{fig:MHDBernoulli}
\end{figure}

We choose the $\gamma_b\alpha_t=200$ simulation presented in Section 4.2.1 and 
plot the 1D profile of $(e_k+e_{B_{\rm tor}}/2)$ along the jet radius and the 
cross-sectional 2D view of $(e_k+e_{B_{\rm tor}}/2)$ and density/flux in Fig. \ref{fig:MHDBernoulli}. The three plots directly illustrate that at any given time after jet collimation is completed, 
$(e_k+e_{B_{\rm tor}}/2)$ is constant on the boundary of a magnetic tower jet through the entire jet body.

Having cross-checked the jet velocity dependence on poloidal current and density 
using experiments, simulation and analytical theory, and also demonstrated that 
Eq. (\ref{eqn:modified_MHD_Bernoulli}) holds along the jet in the simulation, 
we conclude that Eq. (\ref{eqn:jet_velocity}), and more generally, the 
MHD Bernoulli Eq. (\ref{eqn:MHDBernoulli}) are true for magnetic tower jets, 
such as the Caltech experimental plasma jet and possibly actual astrophysical jets.

\section{Sensitivity to Imposed Simulation Conditions}
The numerical simulations presented in Section 4 are based on a number of imposed conditions, including initial mass distribution, background pressure, initial poloidal field, toroidal field injection rate and toroidal field injection volume (factor $A$). As discussed in Section 3 and 4, the initial poloidal field flux and toroidal field injection rate are selected strictly on the experiment properties. The initial mass distribution in simulation is similar to the real experiment case. We now examine how our key conclusions depend on these imposed conditions.

\begin{figure}
\figurenum{15}
\epsscale{1}
\plotone{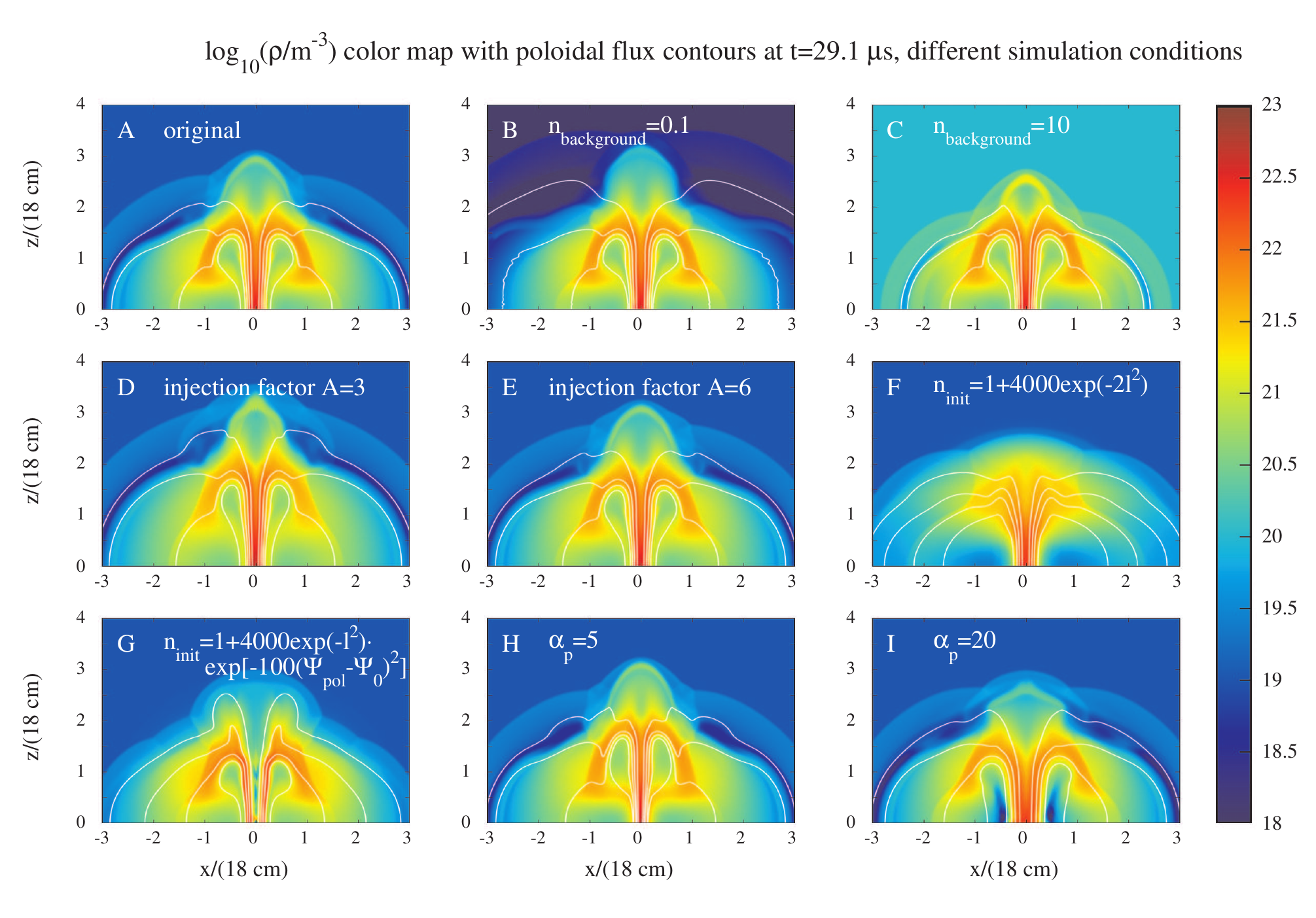} 
\caption{Cross-sectional view of density distribution (color map) and azimuthally-averaged poloidal flux contours (white curves) in $xz$ plane ($z>0$) at $t=0.5$ ($29.1$ $\mu$s) of nine simulations with different conditions. Each plot is formatted the same way as Fig. \ref{fig:den_evolution} except the density range is from $10^{18}$ m$^{-3}$ to $10^{23}$ m$^{-3}$. A: the original simulation described in Section 4.1 with initial mass distribution $n_{\rm init}=n_{\rm background}+4000e^{-l^2}e^{-40[(r-1/2)^2+z^2-1/4]^2}$, background density $n_{\rm background}=1$ (corresponding to $10^{19}$ m$^{-3}$), injection factor $A=9$ and total poloidal flux factor $\alpha_p=10$ (corresponding to a total flux $1.59$ mWb). Panel B-I show simulations with same conditions as simulation A except only \textbf{one} different condition. B: simulation with initial background density $n_{\rm background}=0.1$ ($10^{18}$ m$^{-3}$), $10$ times lower than simulation A. C: simulation with initial background density $n_{\rm background}=10$ ($10^{20}$ m$^{-3}$), $10$ times denser than simulation A. D: simulation with injection factor $A=3$. E: simulation with injection factor $A=6$. F: simulation with initial mass distribution $n_{\rm init}=1+4000e^{-2l^2}$. G: simulation with initial mass distribution $n_{\rm init}=1+4000e^{-l^2}e^{-100(\Psi_{\rm pol}(r,z)-\Psi_0)^2}$. H: simulation with initial poloidal flux factor $\alpha_p=5$ (corresponding to a total flux $0.79$ mWb, $50\%$ of simulation A). I: simulation with initial poloidal flux factor $\alpha_p=20$ (corresponding to a total flux $3.17$ mWb, twice of simulation A). The injection rates $\gamma_b\alpha_t$ of simulation H and I are adjusted correspondingly so that the effective toroidal injection rate $\gamma_b\alpha_t\alpha_p$ of these two simulations are the same with simulation A. Panel A-G are overlaid by poloidal flux contours from $0.2$ mWb to $1.4$ mWb every $0.4$ mWb. Panel H has contours from $0.1$ to $0.7$ mWb every $0.2$ mWb; Panel I has contours from $0.4$ mWb to $2.8$ mWb every $0.8$ mWb.}\label{fig:conditions}
\end{figure}

We perform another eight simulations with exactly the same conditions as the simulation presented in Section 4.1 (referred as the ``original'' simulation or simulation A in the following discussion), except for one different condition. The density distribution and poloidal field configuration at $t=29.1$ $\mu$s of these eight simulations are plotted in Fig. \ref{fig:conditions} together with the original simulation.

\subsection{Background Condition}
The original simulation has a background plasma particle number density $n_{\rm background}=1$, or $10^{19}$ m$^{-3}$, about $10^3-10^4$ times less dense than the central jet (panel A in Fig. \ref{fig:conditions}). In the experiment, this number is $10^7-10^8$. However, as long as the background density is significantly lower than the plasma of interest, the dynamics of the central jet should not be affected.

This is verified by simulation B and C, which have $n_{\rm background}=0.1$ and $10$, respectively. Comparing A, B and C, they show no difference in the central jet and the vicinity. The hydro shock and return flux at very large radii, however, are indeed affected by the different background conditions. Consistent with the discussion in Section 3.2.2, Section 4.1.2 and Section 4.1.3, a lower background pressure imposes a weaker restriction to the expansion of the system.

In an astrophysics situation, the density difference between the central jet and ambient environment (ISM/IGM) is expected to be less than in the experiment and the shock structure and the return flux are expected to be somewhat different. With a significant background pressure, the expansion of return flux and current can be highly constrained. If the return flux and current are sufficiently near the center jet, they can influence the jet stability properties. This is similar to how a conducting wall surrounding a current-carrying plasma tube can prevent the plasma against from developing a kink instability \citep[e.g., ][]{Bellan_Plasma}. 

\subsection{Toroidal Field Injection Condition}
The toroidal field injection condition is subjected to two major possible variations: injection rate and injection volume.

The injection rate affects the total poloidal current and therefore affects the jet velocity according to Eqn. \ref{eqn:jet_velocity}. In Section 4.2, we have addressed this issue by performing five simulations with different injection rates. Figure \ref{fig:jet_velocity} shows that jet velocity is proportional to the toroidal injection rate.

Injection volume is determined by the injection factor $A$ (Section 3.3.1). We already pointed out that the factor $A$ does not alter the total poloidal current associated with the toroidal field. Simulation D and E shown in Fig. \ref{fig:conditions} are performed with $A=3$ and $A=6$, respectively. At $z=1$, the factor $e^{-Az^2}=0.05$, $2.5\times10^{-3}$ and $1.2\times10^{-4}$ for $A=3$ (D), $6$ (E) and $9$ (A), respectively. Even with such enormous differences, the plasmas in simulation A, D and E evolve in very similar ways. This is because the injected toroidal field is able to emerge into the propagating jet rapidly, no matter where the field is initially injected (see also in Fig. \ref{fig:axial_profile}, \ref{fig:energy_flux} and \ref{fig:jet_structure}).

A notable difference for different $A$ factors is the behavior of the hydro shock and remote return flux. This is because toroidal injection with a smaller $A$ gives larger direct field injection at larger distance and low density region, and therefore gives rise to a faster expanding shock and return flux.

The $A$ factor determines the thickness of the effective engine region. In the experiment and astrophysics cases, the engine region is expected to be limited to the electrodes or the vicinity of central objects. Ideally, a toroidal injection with a larger $A$ factor provides better approximation to the real cases. However, the $A$ factor has little effect on the dynamics of the central jet.

\subsection{Initial Mass Distribution}
As shown in Section 3 and Section 4, the jet is created as a result of a gradient along the $z$ direction of the pressure associated with the toroidal magnetic field. Therefore the initial mass distribution should not be crucial in the jet dynamics.

Simulation F adopts a very different initial mass distribution $n_{\rm init}=1+4000e^{-2l^2}$, where $l^2=r^2+z^2$. A central jet is created with a similar radius and slower speed. Further investigation shows that the well-collimated portion extends from $z\approx 0.8$ to $1.1$ in the next $6$ $\mu$s. The return flux manages to expand further because of the relative low density at large radii initially. The general jet behaviors are consistent with simulation A.

Simulation G takes an initial mass distribution very similar to the real experiment case, $n_{\rm init}=1+4000e^{-l^2}e^{-\delta(\Psi_{\rm pol}(r,z)-\Psi_0)^2}$ with $\delta=100$ (see Section 3.2.2). The central region is initially filled with low density plasma. In the experiment, fast magnetic reconnection allows the magnetic field to diffuse into the center along with the plasma. However, in ideal MHD theory, reconnection is forbidden. As shown in panel G of Fig. \ref{fig:conditions}, a hollow jet is eventually formed. The axis magnetic field is stronger along the axis than simulation A, because there is no dense plasma in the center helping the poloidal flux against compression of the toroidal pinch. Because the plasma is initially distributed parallel to the poloidal field, simulation G shows a better alignment between plasma and poloidal flux compared to simulation A.

Although the detailed form of initial mass distribution does not significantly affect the formation of the central magnetic tower jet, it can at later times impact the density distribution at larger radius, such as return flux region, and therefore can potentially influence the expansion of the return current. Three additional simulations A2, A3 and A4 are performed which are the same as simulation A (original one) except that there is less dense plasma at either larger radius or larger height. Table \ref{tab:return_current} lists the detailed function of initial mass distribution and the location of return current at $z=20$ cm for each simulation. Max $B_{\theta}$ in Table \ref{tab:return_current} is the toroidal field strength at the central jet surface. The return flux region ($T$-shell) of A2-A4 is less dense than that of simulation A. This is because initially there was less dense plasma at larger radius or height. As expected, the return current of A2-A4 expands faster than does simulation A. With a lower background pressure, simulation B also has a faster expanding return current than A does.

It is found that all these simulations produce similar magnetic/kinetic profiles in the central region, although their return current profiles differ significantly. This is because, according to Ampere's Law, there is no magnetic field generated by the return current at the central jet location. In both the simulation and experiment, there is no boundary condition constraining the radius of zero net current and hence the return current radius can expand from the MHD force. The return flux region of simulation A expands at speed $v_r\approx v_z\approx 6$ km s$^{-1}$ at $t=29.1$ $\mu$s (Fig. \ref{fig:r_profiles}). This is comparable with the Alfv\'en velocity $V_A\sim 15$ km s$^{-1}$ in the diffuse pinch region between the central jet and the return flux/current.

\begin{deluxetable}{ccccc}
\tablewidth{16cm}
\centering
\tablecaption{Location of return current of simulations with different initial density distribution at $z=20$ cm at $t=24.4$ $\mu$s for simulation A, B A2-A4 and $t=27.9$ $\mu$ for simulation F.}
\tablenum{2}
\tablehead{\colhead{Simulation} & \colhead{initial mass distribution} & \colhead{max $B_{\theta}$ (T)} & \colhead{$R_{J_z=0}$ (cm)} & \colhead{$\bar{R}_{J_z<0}$ (cm)}}
\startdata
A & $1+4000 f(r,z)e^{-r^2-z^2}$ & $0.119$ & $7.3$ & $11.7$ \\ 
B & $0.1+4000 f(r,z)e^{-r^2-z^2}$ & $0.107$ & $8.3$ & $11.9$ \\ 
F & $1+4000 e^{-2r^2-2z^2}$ & $0.083$ & $9.8$ & $22.6$ \\ 
A2 & $1+4000 f(r,z)e^{-r^2-4z^2}$ & $0.123$ & $9.2$ & $16.5$ \\ 
A3 & $1+4000 f(r,z)e^{-2r^2-z^2}$ & $0.091$ & $9.0$ & $14.5$ \\ 
A4 & $1+4000 f(r,z)e^{-2r^2-4z^2}$ & $0.100$ & $10.5$ & $21.6$ \\
\enddata
\tablenotetext{Note }{Function $f(r,z)=e^{-40[(r-1/2)^2+z^2-1/4]^2}$.
$R_{J_z=0}$ is the radius where axial current changes sign and $\bar{R}_{J_z<0}$ is the averaged location of the return current, defined as $\bar{R}_{J_z<0}\equiv (\sum r|J_z|^2) /(\sum|J_z|^2$) for all negative $J_z$. The numbers for simulation F are obtained at $t=27.9$ $\mu$s when the jet has a similar height as other simulation jet at $t=24.4$ $\mu$s (see Fig. 15 panel A and F).}
\label{tab:return_current}
\end{deluxetable}

\subsection{Initial Poloidal Flux}
Compression of the poloidal flux tends to oppose the pinching force of the toroidal field. Simulation H and I verify this with $50\%$ and $200\%$ initial poloidal flux compared to simulation A. Panel H and I of Fig. \ref{fig:conditions} show that with less poloidal flux, the jet has a smaller radius and propagates faster; with doubled poloidal flux, on the other hand, the plasma struggles to compress the poloidal field, resulting in a much wider and slower jet.

In non-axisymmetric situations, the poloidal flux is expected to impact the stability properties of the current-conducting jet. Experiment investigation involving changing the ratio between poloidal current and poloidal flux, known as ``gun parameter'', shows that the jet undergoes MHD kink instability when the classical Kruskal-Shafranov threshold is satisfied
\citep{Hsu_Bellan_2003,Hsu_Bellan_2005}.\\

In summary, we have shown here how different conditions affect the simulation results. The conditions that directly determine the jet dynamics, such as initial poloidal flux and toroidal injection rate, are selected strictly from the actual experiment conditions. Those conditions that only affect the dynamics of return flux and the hydro shock, such as background pressure, initial mass distribution and toroidal injection volume, can be subject to relatively large variations without significantly influencing the jet dynamics.

\section{Summary and Discussion}
We have presented MHD numerical simulations of the Caltech plasma jet experiment 
using a magnetic tower model similar to \citet{Li2006_MagneticTower}. 
By having a purely toroidal magnetic injection localized around the $z=0$ plane, 
the simulation jet gains energy and helicity in a manner analogous to the 
electrode-driven experimental jet, or to astrophysical jets driven by accretion disks. 
In the simulation, the injected toroidal field near $z=0$ is efficiently carried through 
the jet and is responsible for generating the pinch force that collimates both the 
plasma and the embedded poloidal magnetic field. The gradient of the collimation force 
along the jet boundary, or equivalently, the gradient of toroidal magnetic field energy in the $z$ direction, 
is responsible for accelerating the jet.
Magnetic to kinetic energy conversion is 
verified in the simulation along with the experiment.

The simulation jet agrees quantitatively with the experimental jet in numerous ways, 
including the energy partition/evolution, current/voltage, jet radius, axial profile, 
magnetic field structure and jet velocity scaling. Furthermore, by using the unit systems 
given in Table \ref{tab:units}, the simulation results can easily be made dimensionless 
and then converted to astrophysical scales.

One of the most significant outcomes of this simulation work is the validation of using 
terrestrial laboratory experiments to study astrophysical jets. Although it is not feasible to experimentally reproduce every single aspect of an astrophysical jet, by careful experiment design it is possible to replicate many of the most important mechanisms that govern the jet dynamics. 
Also, the experimental 
investigation shares common advantages with the numerical simulation such as 
reproducibility, freedom in parameter space and possibility of in-situ measurement. 
This paper suggests that combining observation, theoretical modeling and laboratory 
experiments helps understand the nature of magnetically driven plasma flows.

We emphasize here that the simulation does not prove that the experimental jets are exactly the same as astrophysical jets. Neither the simulation nor the experiment is expected to reproduce every detail of a theoretical model or an astrophysical jet. However, the fact that an astrophysical magnetic tower model can be used to simulate laboratory experiments suggests that the experiment shares several important similarities with astrophysical jets, such as the collimation and propagation mechanisms. Furthermore consideration of any discrepancies between experiment and simulation help understand the underlying physics.

In both the experiment and simulation, there is no boundary condition or other restriction on the expansion of the return current/flux. The return current/flux expands at a velocity comparable to Alfv\'enic velocity but the dynamics of the central magnetic tower jet is not influenced by the return current/flux. In astrophysical situations where the background pressure is important \citep[e.g., ][]{Lynden-Bell_1996,Li_2001}, free expansion of the return current/flux can be inhibited, resulting in a small or null diffuse pinch region, i.e., the return current/flux could be snugly on the surface of the central jet \citep[e.g. ][]{SherwinLyndenBell_2007, Nakamura_2007}. In this situation, most of the toroidal field energy is inside the central jet so the jet is expected to be more efficiently collimated and accelerated for a fixed amount of toroidal energy. Meanwhile, an extremely dense return flux region closed to the jet could act like a wall that would stabilize the central jet.

The simulation presented in this paper mainly addresses jet launching and acceleration mechanisms, 
i.e., jet collimation, propagation and energy conversion, and considers only axisymmetric 
dynamics. No asymmetric perturbation is introduced initially or during the simulation. 
The simulation jets, theoretically vulnerable to kink instability, remain quasi-axisymmetric 
and stable. However, preliminary investigation has been able to produce kink instability 
in the simulation, by using a perturbed initial mass distribution. In the experiment, 
due to the inevitable imperfectly symmetric laboratory condition, the jet always undergoes 
kink instability when the classic Kruskal-Shafranov condition is satisfied 
\citep{Hsu_Bellan_2003,Hsu_Bellan_2005}. In some cases when the kinked plasma 
grows exponentially fast and accelerates away from the central axis, a lateral 
Rayleigh-Taylor instability is induced on the inner boundary of the jet. 
The Rayleigh-Taylor instability further induces a fast magnetic reconnection that 
breaks the jet in the middle, and removes some magnetized jet segment from the 
electrode-attached jet segment \citep{Moser_Nature}. Astrophysical jets in a similar situation, 
e.g., kink instability or other lateral acceleration, might also be susceptible 
to this secondary instability. Numerical investigation of this Rayleigh-Taylor 
instability is underway.


\acknowledgments
The experimental program at Caltech is supported by the NSF/DOE Partnership in Plasma Science.
H. L. is grateful to Stirling Colgate, Ken Fowler, and Ellen Zweibel for discussions. 
H. L. and S. L. are supported by the LANL/LDRD and Institutional Computing Programs at 
LANL and by DOE/Office of Fusion Energy Science through CMSO.

\end{document}